\def\wide{{\rm wide}}
\def\narr{{\rm narr}}

\def\beq{\begin{eqnarray}}
\def\eeq{\end{eqnarray}}

\def\beqn{\begin{eqnarray*}}  % no equation numbers are generated
\def\eeqn{\end{eqnarray*}}

\def\E{{\rm E}}
\def\Var{{\rm Var}}

\def\dd{{\rm d}}
\def\N{{\rm N}}

\def\Pr{P}
\def\pr{{\rm pr}}

\def\quadandquad{\quad \hbox{and} \quad}
\def\arr{\rightarrow}
\def\hatt{\widehat}

\def\sumin{\sum_{i=1}^n}

\def\eps{\varepsilon}
\def\half{\hbox{$1\over2$}}

\def\rootn{\sqrt{n}}

\def\midd{\,|\,}
\def\tr{{\rm t}}
\def\dell{\partial}

\def\KL{{\rm KL}}

\def\dellone{\hbox{$\dell\mu\over \dell\theta$}}
\def\delltwo{\hbox{$\dell\mu\over \dell\gamma$}}

\def\true{{\rm true}}

\def\FIC{{\rm FIC}}

\def\mse{{\rm mse}}

\def\pois{{\rm Pois}}

\def\fic{{\rm FIC}} %% or fic, if we wish to 
\def\afic{{\rm afic}}

\def\bsq{{\rm bsq}}

\def\yeartt{{\tt year}}
\def\sex{{\tt sex}}
\def\diatom{{\tt diatom}}
\def\datett{{\tt date}}
\def\age{{\tt age}}
\def\region{{\tt region}}
\def\bodylength{{\tt bodylength}}
\def\fatweight{{\tt fatweight}}
\def\twott{{\tt 2}}

\documentclass[utf8]{frontiersSCNS}
\usepackage{color}

\usepackage{url,hyperref,lineno,microtype,subcaption}
\usepackage[onehalfspacing]{setspace}

\def\keyFont{\fontsize{8}{11}\helveticabold }
\def\firstAuthorLast{Claeskens, Cunen, Hjort} % use et al only if is more than 1 author
\def\Authors{Gerda Claeskens$^{1}$, C\'eline Cunen$^{2}$, 
   and Nils Lid Hjort$^{2,*}$}

\def\Address{$^{1}$ORSTAT and Leuven Statistics Research Centre,
   KU Leuven, Belgium \\
$^{2}$Department of Mathematics, University of Oslo, Norway}

\def\corrAuthor{Nils Lid Hjort}
\def\corrEmail{nils@math.uio.no}

\begin{document}
\onecolumn
\firstpage{1}

\title[Focused Information Criteria]{Model Selection via Focused Information Criteria
   for Complex Data in Ecology and Evolution}

\author[\firstAuthorLast ]{\Authors} % This field will be automatically populated
\address{} % This field will be automatically populated
\correspondance{} % This field will be automatically populated

\extraAuth{}% If there are more than 1 corresponding author,
% comment this line and uncomment the next one.
% \extraAuth{corresponding Author2 \\ Laboratory X2, Institute X2,
% Department X2, Organization X2, Street X2, City X2 , State XX2
% (only USA, Canada and Australia), Zip Code2, X2 Country X2, email2@uni2.edu}

\maketitle

\begin{abstract}
\noindent
Datasets encountered when examining deeper issues
in ecology and evolution are often complex.
This calls for careful strategies for both model building,
model selection, and model averaging.
Our paper aims at motivating, exhibiting, and further developing 
focused model selection criteria. In contexts involving 
precisely formulated interest parameters, these versions of FIC, 
the focused information criterion, typically lead to better final 
precision for the most salient estimates, 
confidence intervals, etc.~as compared to estimators 
obtained from other selection methods.
Our methods are illustrated with real case studies
in ecology; one related to bird species abundance
and another to the decline in body condition for
the Antarctic minke whale.

\tiny
\keyFont{ \section{Keywords:}
bird species abundance,
ecology,
evolution,
FIC and AFIC,
focused model selection,
linear mixed effects,
minke whales}
% All article types: you may provide up to 8 keywords; at least 5 are mandatory.
\end{abstract}

\section{Introduction}
\label{section:intro}

Only rarely will initial modelling efforts lead to
`one and only one model' for the data at hand.
This simple empirical statement applies in particular
to situations with complex data for complicated
and not-yet-understood mechanisms underlying the phenomena
being studied, in ecology and evolution, as well
as other sciences. Thus methods for model comparison,
model selection, and model averaging are called for.
Not surprisingly there must be several such methods,
since the question `what is a good model for my data?'~cannot
be expected to have a simple and clear-cut answer.

There are indeed several model selection schemes
in the statistics literature, with the more famous
ones being the AIC (the Akaike Information Criterion)
and the BIC (the Bayesian Information Criterion);
see \citet{ClaeskensHjort08a} for a general overview.
The AIC and BIC are able to compare and rank
competing models for a given dataset, as long as they
are all parametric. These and yet other methods
work in an `overall modus', in appropriate senses
comparing overall fit with overall complexity,
but they do not take on board {\it the intended use
of the fitted models}. This is where FIC (the Focused
Information Criterion) comes in, along with certain
relatives. The FIC aims at giving the most relevant
model comparison and ranking, and hence also
pointing to the best model, for {\it the given purpose}.
What this given purpose is depends on the
scientific context. Indeed, two research teams
might ask different focused questions, with the
same data and the same list of candidate models,
and we judge it not to be a contradiction in terms
that three focused questions might have three
different best models.

The present article gives an account of FIC and its
relatives, including also certain extensions of
previously published methods. 
We do have models
for ecology and evolution in mind, though it is
clear that the view is broader: we wish to find
good statistical models for complex data, and
can do so, once crucial and context driven questions
are translated to {\it focus parameters}.
%% {\color{blue} ...} 
Our paper's contribution is twofold. 
(i) We aim at introducing the FIC methodology to researchers 
in ecology and evolution. We have therefore strived to 
include relevant examples, along with some R code. 
We also discuss various topics of interest to applied researchers, 
particularly in Section~\ref{section:discussion}. 
In this partly tutorial spirit, various technical details
have been placed in the appendix. 
(ii) Our article also serves as an outlet for a 
somewhat new FIC framework, termed the `fixed wide model framework',
different from the `local asymptotics framework' used 
in the majority of previous publications. 
Details are in Section~\ref{section:fic-new}, 
with material not been presented in this general form before. 
In particular, the extension of this framework 
to generalised linear models is novel.

To help fix ideas and some basic notation, 
we start with a concrete application. 
We use the dataset from \citet{Handetal94} 
regarding counts of the number of bird species on
fourteen areas, vegetation islands, in the Andes mountains with
p\'aramo vegetation. 
%% The full dataset is given in 
%% Section \ref{section:bird-data} in the appendix. 
In addition to the number of bird species $y$, 
there are four covariates recorded for each such 
vegetation island:
$x_1$, the area of the vegetation island in thousands of square kilometers;
$x_2$, the elevation in thousands of meters;
$x_3$, the distance between the area and Ecuador in kilometers;
and $x_4$, the distance from the nearest island in kilometers.

% # from Hand et al. (1994) case #52.
% # For each "island": N, AR, EL, DEc, DNI;
% # number of species; area [in thousands of sq km];
% # elevation [in thousands of m]; distance from Ecuador [in km];
% # distance to nearest island [in km].

{{\tiny 
\begin{verbatim}
  y       x1      x2       x3      x4
  36      0.33    1.26      36     14
  30      0.50    1.17     234     13
  37      2.03    1.06     543     83
  35      0.99    1.90     551     23
  11      0.03    0.46     773     45
  21      2.17    2.00     801     14
  11      0.22    0.70     950     14
  13      0.14    0.74     958      5
  17      0.05    0.61     995     29
  13      0.07    0.66    1065     55
  29      1.80    1.50    1167     35
   4      0.17    0.75    1182     75
  18      0.61    2.28    1238     75
  15      0.07    0.55    1380     35
\end{verbatim}
}}

We model the number of bird species $Y$ by a Poisson distribution with
mean $\exp(x^\tr\beta)$, where $x$ in the widest model
consists of the constant 1 (modelling the intercept),
all four covariates $x_1,\ldots,x_4$ as main effects,
and all six pairwise interactions
between these main effects. This amounts to a total of 11 parameters
$\beta_0,\ldots,\beta_{10}$. We wish to include the 
intercept parameter $\beta_0$ in all candidate models,
and hence take it as a `protected parameter',
whereas the other parameters are `open', and
can be pushed in and out of candidate models. 
% The intercept $\beta_0$ is protected for selection 
% since we wish to include an intercept in all considered models. 
For this application, all submodels of the largest 
11-parameter model are considered, with the further restriction 
that interactions between two covariates can be
included only if the two main effects are present. 
This results in a total of 113 models.

The main distinction between FIC and various 
other information criteria is the presence of a \textit{focus}. 
This is a quantity of interest that depends on the model parameters and is
estimable from the data. The generic notation for the focus 
used in our paper is $\mu$. Its dependence 
on the model parameters might be indicated by writing $\mu(\beta)$.

In the bird species study, our first focus concerns 
%% {\color{blue}} 
one of the vegetation islands, Chiles. 
This area is the one among the fourteen that is 
closest to Ecuador, and has covariate values
$x_1=0.33$, $x_2=1.26$, $x_3=36$, $x_4=14$.
We wish to select a model that best estimates the expected number of
bird species for this island, that is, 
$\mu(\beta) = \exp(x^\tr\beta)$ for the given covariate values for Chiles.
In our model search problem there are 113 models and 
hence 113 estimators for $\mu$. Each such estimator, 
say $\hatt\mu_M$ for a candidate model $M$, 
comes with its own bias and variance, say $b_M$ and $\tau_M^2$.
Thus, for each candidate model there is a corresponding 
mean squared error (mse)
\beq
\label{eq:mseM}
\mse_M=\tau_M^2+b_M^2.
\eeq
The basic idea of the FIC is to estimate these $\mse$
values from the data, for the wide as well as for each 
candidate model, i.e.~to construct  
\beq
\label{eq:hereisFIC}
\FIC_M=\hatt\mse_M=\hatt\tau_M^2+\hatt\bsq_M, 
\eeq 
with the second term indicating estimation of the 
squared bias $\bsq_M=b_M^2$. 
In the end one selects the model with the smallest estimated mse.
% We denote by $\fic_M$ an estimator for $\mse_M$.
% In various cases the variance terms $\sigma_M^2$
% are easier to estimate than the squared biases $b_M^2$.

For the bird species application, we use FIC for finding 
the best model to estimate the expected number of bird 
species for Chiles. We use the R package \texttt{fic} 
with the following lines of R code, 
where we fit the wide model, specify the focus function, 
the covariate value in which to evaluate this focus, 
and the specific models that we wish to search through. 
In this example we restrict the built-in all subsets specification 
to only using models that obey the hierarchy principle
(so out of the $2^{10}=1024$ potential submodels, 
only the 113 pointed to above are included). 
\begin{verbatim}
library(fic)
wide.birds = glm(y ~ .^2, data=birds, family=poisson)
focus1 = function(par, X) exp(X %*% par)
inds0 = c(1,rep(0,10)) # only the intercept is in the narrow model
A = all_inds(wide.birds, inds0)  # use all subsets of the wide model
#exclude models with interactions that do not have both main effects:
inds <- with(A,A[!(  A[,2]==0 & (A[,6]==1|A[,7]==1|A[,8]==1) |
   A[,3]==0 & (A[,6]==1|A[,9]==1|A[,10]==1) |
   A[,4]==0 & (A[,7]==1|A[,9]==1|A[,11]==1) |
   A[,5]==0 & (A[,8]==1|A[,10]==1|A[,11]==1)), ])
# specify the X used to evaluate the focus function:
XChiles=model.matrix(wide.birds)[1, ]
fic(wide=wide.birds, inds=inds, inds0=inds0, focus=focus1, X=XChiles)
\end{verbatim}

For each of the 113 models we get via the output 
values of the focus estimate, the estimated bias, standard error, 
and actually two versions of the FIC of (\ref{eq:hereisFIC}),
corresponding to two related but different ways of 
estimating the $b_M^2$ part 
(for details, see Section \ref{section:fic-general}). 
For FIC tables and FIC plots we prefer working with 
the square-root of the FIC, i.e.~estimates of the 
root-mse (rmse) rather than of the mse, as these 
are on the original scale of the focus and easier to interpret.

Table~\ref{Table:birdsoutput}  is constructed from the output for 
a selection of models, including the narrow model (1)
which has a relatively large (in absolute value) bias estimate of 
$-19.035$, a relatively small standard error of 2.247 
and a focus estimate of 20.71; the wide model (113) 
with zero as the bias estimate though with a large standard 
error of 6.051.
This is a typical output: the wide model contains 11 parameters 
to estimate which causes the standard error to be large, 
the narrow model only contains the intercept resulting in 
a small standard error. For the bias estimate the scenario 
is reversed: the wide model has the smallest bias, 
while the narrow model has a larger bias. The balancing 
act of the FIC via the mean squared error finds a compromise.
The selected model (5) results in the smallest value of 
the square root of the estimated mean squared error (rmse). 
Its indicator sequence 10010,000000, with a one for $\beta_0$
and $\beta_3$, and zeroes for the interactions, 
points towards the selected focus $\mu(\beta)=\exp(\beta_0+\beta_3x_3)$ 
with corresponding estimated focus value 38.88. Using the 
wide model would have resulted in a close 38.27 though 
with a larger estimated root mean squared error. 
The wide model only ranks at the 73rd place according to estimated rmse.
Model (20) is selected by the Bayesian information criterion BIC,
it consists of the intercept, all four main effects and 
the interaction between $x_1$ and $x_2$. In the rmse ranking 
it comes at the 42nd place. Model (67) is the one selected 
by the Akaike information criterion, next to the intercept and all
main effects it consists of the interactions 
$x_1x_3$, $x_2x_3$, $x_2x_4$. This models ranks 32nd.

\begin{table}[!ht]
\caption{Bird species example. This table is constructed 
from output of the R function fic for six of the 113 models, 
together with the AIC and BIC values. FIC selection takes 
place via the square root of the estimated mean squared error 
of the focus estimator.
\label{Table:birdsoutput}}
\centering
\begin{tabular}{clrrrrrrrr}
model & coef.~indicators & focus & bias & se & $\sqrt{\FIC}$ & AIC & BIC\\ [0.1cm]\hline \\
1& 10000,000000 &20.714 & $-$19.035 & 2.247 & 19.167 & 143.26 & 143.90 \\
5& \underline{10010,000000} & \underline{38.882} & 0.000 & 4.383 & \underline{4.383} & 112.65 & 113.93
\\
20& 11111,100000 & 33.718 & $-$2.156 & 4.670 & 5.143 & 91.91 & \underline{95.74}
\\
28& 11101,001000 & 26.356 & $-$11.0468 & 3.674 & 11.642 & 98.54 & 101.74
\\
67& 11111,010110 & 39.784 & 0.000 & 5.296 & 5.296 & \underline{91.44} & 96.55
\\
113 &11111,111111 & 38.269 & 0.000 & 6.051 & 6.051 & 95.72 & 102.75
\end{tabular}
\end{table}

The second focus concerns the probability of having more than
30 bird species, $\Pr(Y>30\midd x)$. 
Now we do not specify a particular island 
but use the average FIC (see Section~\ref{subsection:afic}),  
with equal weights for the fourteen vegetation islands 
(non-equal weights can easily be worked with too). 
\begin{verbatim}
focus2 = function(par, X) 1-ppois(30,lambda=exp(X %*% par))
Xall = model.matrix(wide.birds)
fic2 = fic(wide=wide.birds,inds=inds,inds0=inds0,focus=focus2,X=Xall)
AVE = fic2[fic2$vals=="ave",]
which.min(AVE$rmse.adj)
\end{verbatim}
The AFIC selects the following form for the mean:
$\exp(\beta_0+\beta_1x_1 + \beta_2x_2 +\beta_4 x_4+ \beta_7x_1 x_4)$.
The averaged focus estimate of the probability of observing
over 30 bird species in the selected model equals 15.73\%,
while the wide model's estimate is 21.83\%,
though with a substantial larger estimated mean squared error
due to the estimation of 11 parameters instead of only 5
for the selected model.
Of course, AIC and BIC ignore any information regarding the focus,
and thus still recommend the very same models, 
model (67) for AIC, with estimate 21.15\%, 
and model (20) for BIC, with estimate 21.59\%. 
The AIC model ranks 16th, the BIC model is now at the third place.

\begin{figure}[!h]
\begin{tabular}{cc}
(a) & (b) \\[-1.3cm]
\includegraphics[width=0.5\textwidth]{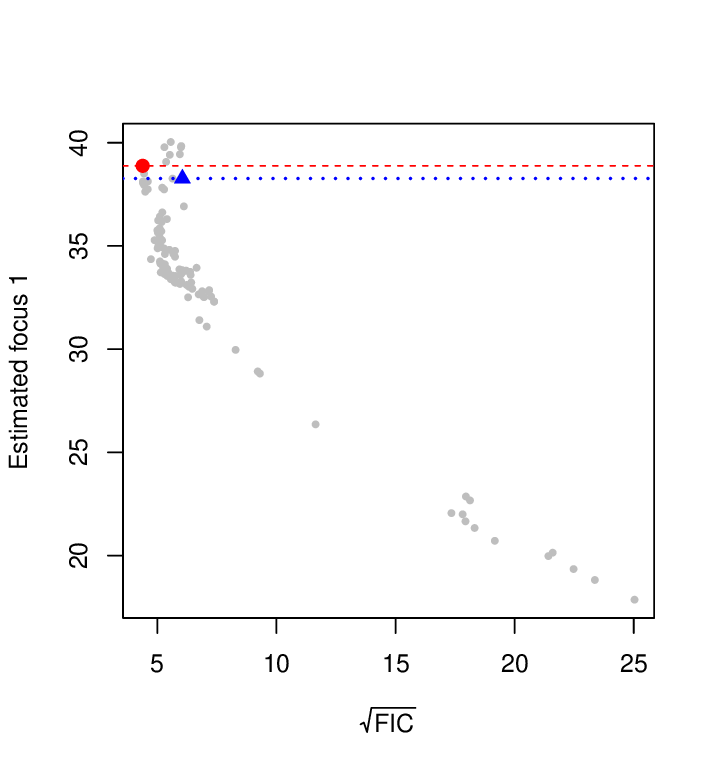} &
\includegraphics[width=0.5\textwidth]{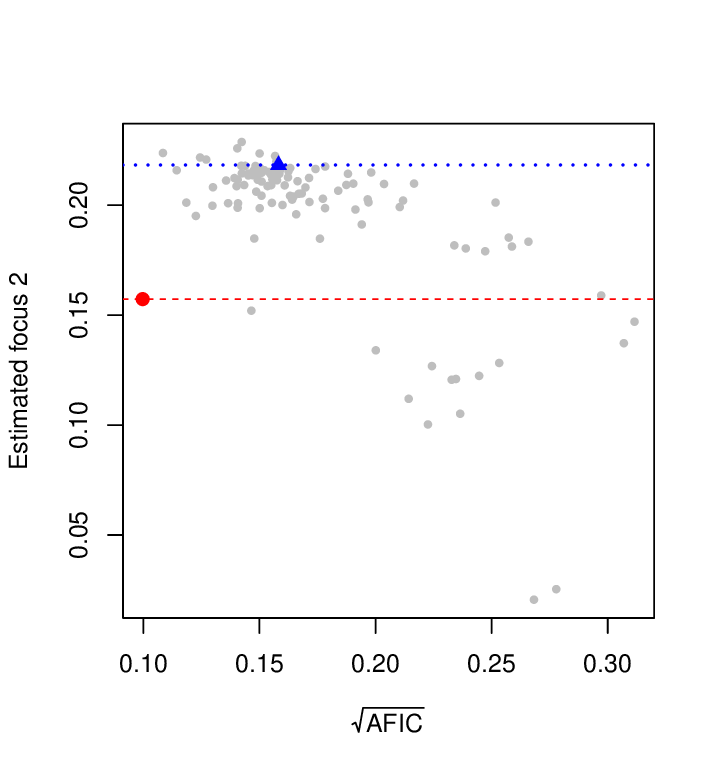}
\end{tabular}
\caption{\label{figure:birds}
The two plots give values for a total of 113 Poisson regression
models, related to two different focused questions.
(a) FIC plot for estimating the expected
number of bird species for the Chiles region.
(b) AFIC plot for estimating the probability of
observing over 30 species, averaging over
all 14 islands. The red dot and line indicate the selected value,
the blue triangle and line are for the wide model.}
\end{figure}

Figure~\ref{figure:birds} displays for these two foci
the root-FIC and root-AFIC values, as well as the estimated
focus values, for all of the 113 models.
The FIC or AFIC selected values, minimising
the respective criteria, are indicated in red,
while the wide model's values are in blue.

Several traditional model selection criteria,
such as the AIC and the BIC
\citep[see][Chs.~2, 3]{ClaeskensHjort08a}
work in an overall modus, finding models that
in a statistical sense are good on average,
not taking on board the specific aims of a study.
The FIC works explicitly with such specific aims,
formalised via the focus parameters. Thus FIC
might find that one model works very well for
covariates `in the middle', whereas another
model could work rather better for covariates
outside mainstream. Similarly, one model might
work well for explaining means, and another
for explaining variances. We stress that the
FIC apparatus works for {\it any specified focus parameter},
and is not limited to e.g.~regression coefficients
and the customary selection of covariates
from that perspective.

The generic FIC formula (\ref{eq:hereisFIC}) cannot be 
immediately applied, as efforts are required to 
establish formulae for approximations to biases and variances, 
along with construction of estimators for these 
quantities. Thus the FIC formula pans out differently 
in different situations, depending on the general framework,
the complexity of models, and estimators of the focus 
parameters. A brief overview of general principles,
leading to such approximations and estimators, 
is given in Section~\ref{section:fic-general}. 
This also encompasses AFIC, ways of creating 
average-FIC scores in situations where 
more than one focus parameter is at stake. 

In Section~\ref{section:fic-new} we provide 
the general FIC formulae in the so-called fixed wide model framework.
The development of FIC formulae ingredients 
in a somewhat different framework, with 
local neigbourhood models, is placed in 
Section~\ref{section:details}. 
Generalised linear models are used as examples, encompassing
linear regression, logistic and Poisson regression, etc.
The more general class of linear mixed models
has proven important for various applications
to ecology, and in Section~\ref{section:fic-lme}
FIC formulae are reached for such.
In Section~\ref{section:whales}
we use linear mixed effects models with FIC 
for analysing the body conditions of minke whales
in the Antarctic, where one focus parameter 
is the yearly decline in energy storage. 
A general but brief discussion is then offered 
in Section~\ref{section:discussion}. Here we touch 
on aspects of performance, along with a few 
concluding remarks, some of which point to future research.

\section{Focused information criteria}
\label{section:fic-general}

The application concerning birds on vegetation islands 
in the previous section was meant to provide intuition 
for the use of FIC for model selection. Here we give 
a more formal, but brief, overview of the FIC and AFIC schemes.
%% [xx check all 1/n things. xx]
%More formalism and notation than in the example. 

\subsection{General FIC scheme}
\label{subsection:fic}

% In the following, we will present a model selection 
% situation which involves the choice between only two models, the 

Suppose we have defined a {\it wide model} which is assumed 
to be the true data-generating mechanism. Estimating the 
focus parameter using the wide model leads to $\hatt\mu_\wide$,
which under broad regularity conditions will aim at 
%% [xx tend to xx]
$\mu_\true$, the unknown true value of the focus parameter. 
Estimation via fitting a candidate model $M$ leads to
$\hatt\mu_M$, say, aiming for some least false
parameter $\mu_{0,M}$, typically different from $\mu_\true$,
due to modelling bias. The least false parameter
in question relates to the best approximation
candidate model $M$ can manage to be, to the true model.
There is therefore an inherent bias, say
\beqn
% \label{eq:bb}
b_M=\mu_{0,M}-\mu_\true,
\eeqn
associated with using $M$. We saw estimates of this 
bias in the birds application above, 
where small models could have larger biases.

The estimators will have certain variances. 
%% cf.~equations (\ref{eq:mseM})--(\ref{eq:hereisFIC}). 
In most frameworks, involving independent or weakly
dependent data, these tend to zero with speed $1/n$,
in terms of growing sample size $n$. 
It is therefore convenient and informative to write 
these variances as $\tau_\wide^2=\sigma_\wide^2/n$ 
and $\tau_M^2=\sigma_M^2/n$, where the mathematics
and approximation theorems associated with different
frameworks typically yield expressions for or approximations
to the $\sigma_\wide$ and $\sigma_M$. 
The $\mse$ of the focus parameter estimators 
is the sum of the variance and the bias squared,
\beq
\label{eq:msemse}
\mse_\wide=\sigma_\wide^2/n+0^2
   \quadandquad
\mse_M=\sigma_M^2/n+b_M^2.
\eeq
These quantities are measures of the \textit{risk}, 
in the statistical sense, associated with using each of 
the models for estimating $\mu$. As explained in the introductions, 
the FIC scores of (\ref{eq:hereisFIC}) are estimates 
of the $\mse$ of the focus parameter estimators, 
i.e.~the $\hatt\mu_M$, for a specific dataset, 
for each of the models under consideration.
Eq.~(\ref{eq:msemse}) is also an informative reminder 
that with more data, variances get small, but biases remain.
So using a model which is not fully correct can still
yield sharper estimators, as long as the bias is moderate
or small: $|b_M|<(\sigma_\wide^2-\sigma_M^2)^{1/2}/\rootn$. 
It is also clear that with steadily more data, 
steadily more sophisticated models can and indeed should 
be used. The FIC makes these ideas operative.  

In various cases the variance terms $\sigma_M^2/n$
are easier to estimate than the squared biases $b_M^2$.
A starting point for the latter is 
$\hatt b_M=\hatt\mu_M-\hatt\mu_\wide$,
but the corresponding $\hatt b_M^2$ will
%, due to well-known relations between expectations and variance, 
overshoot $b_M^2$ with about $\kappa_M^2/n$, 
which is the variance of $\hatt b_M$.
With appropriately constructed estimators of the quantities 
$\sigma_\wide$, $\sigma_M$, $\kappa_M$ 
(with different recipes for different situations), 
this yields two  natural ways of estimating 
the actual $\mse$ values: 
\beq
\label{eq:twofic}
\begin{array}{rcl}
\fic_\wide^u&=&\displaystyle 
\hatt\sigma_\wide^2/n+0^2
   \quadandquad
\fic_M^u=\hatt\sigma_M^2/n+\hatt b_M^2-\hatt\kappa_M^2/n, \\
\fic_\wide&=&\displaystyle 
\hatt\sigma_\wide^2/n+0^2
   \quadandquad
\fic_M=\hatt\sigma_M^2/n+\max(\hatt b_M^2-\hatt\kappa_M^2/n,0).
\end{array}
\eeq
The $\fic^u$ scores are (approximately) unbiased estimates
of the $\mse$, since $\hatt b_M^2-\hatt\kappa_M^2/n$
is (approximately) unbiased for $b_M^2$, 
whereas the $\fic$ scores are adjusted versions,
by truncating any negative estimates of squared bias 
to zero, as we did in the first example.
If the true bias in question 
is some distance away from zero, $\fic_M^u$ will be equal to $\fic_M$.
When faced with a specific application one should decide 
on one of these two FIC versions, and use the same choice 
for all models under consideration.
% At any rate, the model ranking and selection recipe 
% consists in computing all $\fic_M^u$ or $\fic_M$ scores, 
% and trust models with smaller FIC values the most.
% [xx can use either version, but for each analysis be 
% consistent and use the same type for each model xx]

In order to turn the general scheme (\ref{eq:twofic})
into clear formulae, with consequent algorithms,
we need expressions for or approximations to
the population quantities $\sigma_M$, $b_M$, $\kappa_M$,
followed by clear estimation strategies for these again.
In most cases we need to rely on large-sample
approximations. Arriving at clear formulae for
$\sigma_M$ etc.~depends on the particularities of
the wide model, the candidate models, and the focus
parameter. 
We provide such FIC formulae, for
{\it two different frameworks} or setups.
{\it The first} involves local asymptotics,
with candidate models being a local distance
$O(1/\rootn)$ away from the wide model. 
%% where $n$ denotes the sample size.
This derivation is placed in Section~\ref{section:details}.
{\it The second} avoids such local asymptotics and works from
a fixed wide model and a collection of candidate models, 
see Section~\ref{section:fic-new}.
It is not a contradiction in terms that these
two frameworks lead to related but not identical
FIC formulae, as different mathematical approximations
are at work.

\subsection{AFIC, the averaged-weighted selection scheme}
\label{subsection:afic}

%% check all 1/n things 

The FIC apparatus above is tailored to one specific
focus parameter at a time. In a regression context
this applies e.g.~to estimating the mean response
function for one covariate vector at a time,
say $\mu(\theta;x_0)$. Often there would be active interest
in several parameters, however, as with such
a $\mu(\theta;x_0)$ for all $x_0$ in a segment of
covariates, or a probability $\Pr(Y\ge y_0\midd x_0)$
for a set of thresholds, as in the birds study.  

Suppose in general that an ensemble of estimands is
of interest, say $\mu(\theta;v)$ with $v\in V$, and that
a measure of relative importance $\dd W(v)$ is assigned
to these. There could be only a few such estimands  under
scrutiny, say $\mu_j$ for $j=1,\ldots,k$,
along with weights of importance $w_1,\ldots,w_k$.
Estimation involving all higher quantiles,
or all covariates within a certain region,
however, would constitute examples where we need
the more general $v\in V$ notation.
Here we sketch the AFIC approach, for estimating
the relevant integrated weighted risk.

% Starting with the fixed wide model framework,
% for each estimand the machinery of Section \ref{subsection:fic-ii}
% applies, leading to
For each focus parameter in the ensemble of estimands 
there is an associated $\mse$ or risk, $\mse(v)$. 
The combined risk associated with using model $M$ then becomes 
% $$\E\,\{\hatt\mu_M(v)-\mu_\true(v)\}^2
%   \doteq \sigma_M(v)^2/n+b_M(v)^2, $$
% with the appropriate $\sigma_M(v)$
% and $b_M(v)=\mu_{0,M,n}(v)-\mu_\true(v)$.
% The combined loss associated with using model $M$ is
% $\int\{\hatt\mu_M(v)-\mu_\true(v)\}^2\,\dd W(v)$,
% and the implied risk becomes
\beqn
r_n(M) %=\E\,L_{n,M}
   =\int \mse(v)\,\dd W(v)
   = \int \{\sigma_M(v)^2/n+b_M(v)^2\}\,\dd W(v), 
\eeqn
with the appropriate $\sigma_M(v)$ and 
$b_M(v)=\mu_{0,M,n}(v)-\mu_\true(v)$.
% The AFIC scores are estimates of this combined risk.
An approximately unbiased estimate of this combined risk is
$$\afic^u(M)
   =\int\{\hatt\sigma_M(v)^2/n\}\,\dd W(v)
   +\int\{\hatt b_M(v)^2-\hatt\kappa_M(v)^2/n\}\,\dd W(v). $$
This is the same as a direct weighted sum or integral
of the individual $\fic^u(M,v)$ scores. 
The adjusted version, however, where a potentially negative value 
of the estimated integrated squared bias is being truncated
to zero, is not identical to the integral
of the $\fic(M,v)$ scores. It is rather equal to
\beqn
%% \label{eq:afic}
\afic(M)=\int\{\hatt\sigma_M(v)^2/n\}\,\dd W(v)
   +\max\Bigl[\int\{\hatt b_M(v)^2-\hatt\kappa_M(v)^2/n\}\,\dd W(v),0\Bigr].
\eeqn

As with FIC, there are two related, but not identical,
approximation schemes, the fixed wide model setup
and the local asymptotics, of respectively
Section~\ref{subsection:fic-ii} and Section~\ref{subsection:fic-i},
leading now to somewhat different AFIC formulae.
% There is a similar setup leading to AFIC formulae
% for the $O(1/\rootn)$ framework of Section \ref{subsection:fic-i},
% where the task is to estimate an ensemble
% of estimands $\mu(v)=\mu(v,\theta,\gamma)$, for $v\in V$.
% This involves in particular the estimand dependent parameter
% $\omega_n(v)=J_{n,10}J_{n,00}^{-1}{\dell\mu(v)\over \dell\theta}
%   -{\dell\mu(v)\over \dell\gamma}$.
For details and applications,
see \citet[Ch.~6]{ClaeskensHjort08a}, \citet{ClaeskensHjort08b}.

There is a connection between Akaike's information criterion AIC 
and AFIC with certain model dependent weights, see
\citet[Sec.~6.2]{ClaeskensHjort08b}. Broadly speaking, 
the AIC turns out to be large-sample equivalent to 
cases with AFIC where `all things are equally important'. 
%% The AFIC and AIC are in some situations large-sample equivalent.

% For an application of AFIC to the example on bird ecology, 
% see the introduction.

\section{FIC within a fixed wide model framework}
\label{section:fic-new}

The FIC as used in the bird species example is the version as
derived in \citet{ClaeskensHjort03}, see also 
\citet[][Ch.~6]{ClaeskensHjort08a}.
For the estimation of bias and variance a local asymptotic 
framework is used in which the parameters of the true 
density of the data are assumed to be of the form
$\gamma=\gamma_0+\delta/\rootn$, with $n$ the sample size, 
see Section~\ref{section:details} for more explanation. 
This assumptions means in practice that we believe 
that all models are relative close to each other and to the
truth. Moreover, all models are submodels of a wide model.
Since the derivation of the FIC formulae is contained 
in the references above, we only place a summary in the appendix.

In this section we present the 
`fixed wide model' framework, which is particularly useful
if the set of candidate models are seen as not being
in a reasonable vicinity of each other. 
% {\color{blue} [xx Here `fixed' refers to the fact 
% that the wide model does not change with the sample size. xx]}
This second framework allows
candidate models of a different sort from the
wide model; in particular, a candidate model does
not have to be a clear submodel of the wide model.
%% {\color{blue} 
Keep in mind that the two different 
FIC frameworks have the same aims and motivation; 
the difference between them lie in the  different 
mathematical tools for estimating the relevant $\mse$ 
quantities, which lead to different formulae. 
In the discussion section \ref{section:discussion} 
we come back to some differences between 
the two frameworks. 
Here we start in Subsection \ref{subsection:fic-ii} 
by presenting the fixed wide model FIC in a general 
regression setup. Then in the two following 
subsections we deal with two specific model classes 
of general interest, generalised linear models 
and linear mixed models, in more detail.

\subsection{General regression models}
\label{subsection:fic-ii}

In this subsection we use the familiar $(x_i,y_i)$
notation for the regression data,
with $x_i$ the covariate vector in question.
The FIC machinery we develop here
%does not use local asymptotics, with the $O(1/\rootn)$ biases,
starts from the existence of a fixed
wide model. The development represents
an extension of earlier work of
\citet{JullumHjort17, JullumHjort19} for i.i.d.~data and survival analysis,
\citet*{KoHjort19} for copulae models,
\citet*{CunenHjortNygaard19} for power-law
distributions (with applications to war and conflict data)
and \citet*{CunenWalloeHjort19a, CunenWalloeHjort19b}
for linear mixed effects models (with application to whale ecology).

Since we wish to estimate the mse of the focus estimator in different models,
we first consider the asymptotic distribution of the parameter estimator
in the wide model and next in the other models of interest. The distributions
are used to form the mse's of the focus estimators and finally we construct
the fic as an estimated mse and select the model with the smallest fic value.

Suppose a wide model density is agreed upon, of the
form $f(y_i\midd x_i,\theta)$, for a certain
parameter vector $\theta$, of length $p$. We consider 
this to be the true model. This $\theta$ would typically 
encompass both regression coefficients and parameters 
related to the spread and shape of error distributions. With
$u(y_i\midd x_i,\theta)=\dell\log f(y_i\midd x_i,\theta)/\dell\theta$
the score function, and
% \beqn
$J_n=n^{-1}\sumin\Var_\wide\,u(Y_i\midd x_i,\theta_\true)$
% \eeqn
the normalised Fisher information matrix
at the true parameter. 
%% {\color{blue} 
Under mild regularity conditions we have the following 
well-known result for the maximum likelihood estimator 
$\hatt\theta_\wide$,
\beq
\label{eq:approx1}
\rootn(\hatt\theta_\wide-\theta_\true)
   \approx_d\N_p(0,J_n^{-1}).
\eeq 
The notation indicates approximate multinormality 
to the first order as the sample size grows, and can also 
be supplemented with a clear limit distribution statement, 
in that case involving a limit covariance matrix $J$ 
for $J_n$. 
% we have the representation
% \beqn
% \hatt\theta_\wide=\theta_\true+J_n^{-1}\bar U_n+o_\pr(n^{-1/2})
% \eeqn
% for the maximum likelihood estimator $\hatt\theta_\wide$,
% with $\bar U_n$ the average of score functions
% $u(y_i\midd x_i,\theta_\true)$. This also leads to
% \beq
% \label{eq:approx1}
% \rootn(\hatt\theta_\wide-\theta_\true)
%   \approx_d J_n^{-1}\rootn\bar U_n
%   \approx_d\N_p(0,J_n^{-1}),
% \eeq
% along with a precise limit distribution theorem
% under mild ergodic regularity conditions,
% involving then the positive definite limit matrix $J$
% of the $J_n$ as the sample size $n$ increases.
%% 
Consider now a candidate model $M$, different from
the wide one, perhaps also in structure and form.
With notation $f_M(y_i\midd x_i,\theta_M)$ for its density, 
and $u_M(y\midd x_i,\theta_M)$ for its score function, 
we have a maximum likelihood estimator $\hatt\theta_M$,
of length $p_M$, maximising the log-likelihood function
$\ell_{n,M}(\theta_M)=\sumin\log f_M(y_i\midd x_i,\theta_M)$.
If the wide model is considered to be the truth, 
the estimator in model $M$ does not necessarily aim 
at the true parameter, but at the least false parameter
$\theta_{0,M,n}$, which is the minimiser of the 
Kullback--Leibler distance from the data-generating mechanism
to the model; 
see details in Section \ref{subsection:detailsficwide}.
% $$\KL_n(f_\wide,f_M(\cdot,\theta_M))
%   =n^{-1}\sumin\int f(y_i\midd x_i,\theta_\true)
%   \log{f(y_i\midd x_i,\theta_\true)\over f_M(y_i\midd x_i,\theta_M)}\,\dd y_i. $$
% Secondly, here we have
% $$\hatt\theta_M=\theta_{0,M,n}+J_{M,n}^{-1}\bar U_{M,n}+o_\pr(n^{-1/2}), $$
% with $\bar U_{M,n}$ the average of score functions
% $u_M(y_i\midd x_i,\theta_{0,M,n})$ for that model, and with
% $$J_{M,n}=-n^{-1}\sumin\E_\wide\big[\,{\dell^2\log f(Y_i\midd x_i,\theta_{0,M,n})
%   \over \dell\theta_M\,\dell\theta_M^\tr}\big]. $$
The estimator in the candidate model has a limiting 
multinormal distribution, with a sandwich type variance matrix,  
\beq
\label{eq:approx2}
\rootn(\hatt\theta_M-\theta_{0,M,n})
   %\approx_d J_{M,n}^{-1}\rootn\bar U_{M,n}
   \approx_d \N_{p_M}(0,J_{M,n}^{-1}K_{M,n}J_{M,n}^{-1}),
\eeq
where
\beqn
J_{M,n}=-n^{-1}\sumin\E_\wide\,{\dell^2\log f(Y_i\midd x_i,\theta_{0,M,n})
   \over \dell\theta_M\,\dell\theta_M^\tr}
\quadandquad 
K_{M,n}=n^{-1}\sumin\Var_\wide\,u_M(Y_i\midd x_i,\theta_{0,M,n}). 
\eeqn
The variance matrices here are defined with respect to
the wide model, at position $\theta_\true$.

From approximations (\ref{eq:approx1})--(\ref{eq:approx2})
the delta method may be called upon to read off
relevant expressions for the approximate distributions
of the focus parameter estimators 
$\hatt\mu_\wide=\mu(\theta)$ and $\hatt\mu_M=\mu_M(\theta_M)$,
where the latter is aiming for the least false parameter 
value $\mu_{0,M,n}=\mu_M(\theta_{0,M,n})$ associated with model $M$. 
Crucially, we also need a multinormal approximation 
to the {\it joint} distribution of $(\hatt\mu_\wide,\hatt\mu_M)$,
in order to assess the distribution of the bias estimator
$\hatt b_M=\hatt\mu_M-\hatt\mu_\wide$; without that 
part we can't build an appropriate estimator for $b_M^2$. 
In the appendix, Section \ref{subsection:detailsficwide}, 
we go through such arguments, and reach 
\beq
\label{eq:approx3}
\begin{pmatrix}
\rootn(\hatt\mu_\wide-\mu_\true) \\
\rootn(\hatt\mu_M-\mu_{0,M,n})
\end{pmatrix}
   \approx_d\N_2(0,\Sigma_{M,n}).
\eeq
Here the $2\times2$ matrix $\Sigma_{M,n}$ has diagonal terms
$c^\tr J_n^{-1}c$ and $c_{M,n}^\tr J_{M,n}^{-1}K_{M,n}J_{M,n}^{-1}c_{M,n}$,
with gradient vectors
\beqn
c=\dell\mu(\theta_\true)/\dell\theta
   \quadandquad
c_{M,n}=\dell\mu(\theta_{0,M,n})/\dell\theta_M
\eeqn
of lengths $p$ and $p_M$. The off-diagonal term of $\Sigma_{M,n}$
takes the form $c^\tr J_n^{-1}C_{M,n}J_{M,n}^{-1}c_{M,n}$,
with a formula for the required covariance related 
term $C_{M,n}$ in the appendix. 

From (\ref{eq:approx3}) we can read off mse approximations, 
\beqn
\mse_\wide\doteq c^\tr J_n^{-1}c/n+0^2 
   \quadandquad
\mse_M\doteq c_{M,n}^\tr J_{M,n}^{-1}K_{M,n}J_{M,n}^{-1}c_{M,n}+b_M^2, 
\eeqn 
with bias $b_M=\mu_{0,M,n}-\mu_\true$. For the latter we use 
the estimator $\hatt b_M=\hatt\mu_M-\hatt\mu_\wide$, 
where the result above also leads to a clear approximation
for the distribution of $\rootn(\hatt b_M-b_M)$. 
This leads to FIC formulae, unbiased and adjusted, as  
\beq
\label{eq:fictwo}
\begin{array}{rcl}
\fic_\wide^u&=&\displaystyle
\hatt c^\tr\hatt J_n^{-1}\hatt c/n+0^2
   \quadandquad
\fic_M^u=\hatt c_M^\tr\hatt J_M^{-1}\hatt K_M
   \hatt J_M^{-1}\hatt c_M/n
   +\hatt b_M^2-\hatt\kappa_M^2/n, \\
\fic_\wide&=&\displaystyle
\hatt c^\tr\hatt J_n^{-1}\hatt c/n+0^2
   \quadandquad
\fic_M=\hatt c_M^\tr\hatt J_M^{-1}\hatt K_M
   \hatt J_M^{-1}\hatt c_M/n
   +\max(\hatt b_M^2-\hatt\kappa_M^2/n,0).
\end{array}
\eeq
% [xx Section~\ref{section:details} contains the 
% {\color{blue} derivation} of \eqref{eq:fictwo}. 
% a bit unclear this. xx]
% [xx clean up the following carefully; make sure
% quantities worked with are defined; check connection
% to appendix. xx]
Here $\hatt c$ and $\hatt c_M$ emerge by
computing gradients of $\mu(\theta)$ and $\mu_M(\theta_M)$
at their respective maximum likelihood positions,
and $\hatt J_n$, $\hatt J_M$ are computed as
normalised observed Fisher information matrices,
for the wide and for the candidate model in question;
specifically, $\hatt J_M$ is $1/n$ times
minus the Hessian matrix from the log-likelihood,
$-\dell^2\ell_{n,M}(\hatt\theta_M)/\dell\theta_M\dell\theta_M^\tr$.
Also, the $p_M\times p_M$ matrix $\hatt K_M$ is
$n^{-1}\sumin\hatt u_{M,i}\hatt u_{M,i}^\tr$,
with $\hatt u_{M,i}=u_M(y_i\midd x_i,\hatt\theta_M)$.
Finally, the $\hatt\kappa_M^2/n$ estimates involves
also the $p\times p_M$ matrix $\hatt C_M$,
which is $n^{-1}\sumin \hatt u_{\wide,i}\hatt u_{M,i}^\tr$.
Model selection proceeds by computing $\fic_M$, 
the estimated mse of the focus estimator $\hatt\mu_M$, 
for all models $M$ of interest, and then selecting that model 
for which this score is the lowest. 

\subsection{FIC for generalised linear models, with a fixed wide model}
\label{subsection:fic-iiii}

We illustrate this FIC machinery for one
popular class of generalised linear models,
namely the Poisson regression models.
Generalisations to other generalised linear models 
are relatively immediate. 
Suppose therefore that we have count data $y_i$ along
with a covariate vector $x_i$ of length $p$. For the fixed
wide model we take the Poisson regression model with
$y_i\sim\pois(\xi_i)$, with $\xi_i=\exp(x_i^\tr\beta)$
containing all covariate information; in particular,
there is also a true parameter $\beta_\true$ there.
Consider then an alternative candidate model $M$
which instead takes the means to be
$\xi_{M,i}=\exp(x_{M,i}^\tr\beta_M)$, with $x_{M,i}$
of length $p_M$, perhaps a subset of the full $x_i$,
or perhaps with some entirely other pieces of covariate information.
Here the log-densities take the form $-\xi_i+y_i\log\xi_i -\log(y_i!)$,
which means
\beqn
\log f=-\exp(x_i^\tr\beta)+y_ix_i^\tr\beta-\log(y_i!)
   \quadandquad
\log f_M=-\exp(x_{M,i}^\tr\beta_M)+y_ix_{M,i}^\tr\beta_M -\log(y_i!),
\eeqn
for the wide model and the candidate model, along with
score functions
\beqn
u(y_i\midd x_i,\beta)=\{y_i-\exp(x_i^\tr\beta)\}x_i
   \quadandquad
u_M(y_i\midd x_{M,i},\beta_M)=\{y_i-\exp(x_{M,i}^\tr\beta_M)\}x_{M,i}.
\eeqn
From this we deduce
\beqn
J_n&=&n^{-1}\sumin \exp(x_i^\tr\beta_\true)x_ix_i^\tr, \\
J_{M,n}&=&n^{-1}\sumin \exp(x_{M,i}^\tr\beta_{0,M,n})x_{M,i}x_{M,i}^\tr, \\
K_{M,n}&=&n^{-1}\sumin \exp(x_i^\tr\beta_\true)x_{M,i}x_{M,i}^\tr, \\
\eeqn
along with the $p\times p_M$ covariance matrix $C_{M,n}$,
defined as
\beqn
n^{-1}\sumin \E_\wide\,\{Y_i-\exp(x_i^\tr\beta_\true)\}x_i
   \{Y_i-\exp(x_{M,i}^\tr\beta_{0,M,n})\}x_{M,i}^\tr
   =n^{-1}\sumin \exp(x_i^\tr\beta_\true)x_ix_{M,i}^\tr.
\eeqn
% \beqn
% C_{M,n}&=&n^{-1}\sumin \E_\wide\,\{Y_i-\exp(x_i^\tr\beta_\true)\}x_i
%    \{Y_i-\exp(x_{M,i}^\tr\beta_{0,M,n})\}x_{M,i}^\tr \\
% &=&n^{-1}\sumin \exp(x_i^\tr\beta_\true)x_ix_{M,i}^\tr.
% \eeqn
Consistent estimates of these population matrices
are obtained by inserting $\hatt\beta_\wide$ for $\beta_\true$
and $\hatt\beta_M$ for $\beta_{0,M,n}$.

Notably, as long as there is a well-defined
wide Poisson regression model, as assumed here,
the framework is sufficiently flexible and broad
to encompass also non-Poisson candidate models.
Using the FIC apparatus involves working with
log-likelihood functions and score functions
for these alternative models, leading to different
but workable expressions for the matrices
$J_{M,n}$, $K_{M,n}$, $C_{M,n}$ above. The stretched
Poisson models used in \citet[Exercise 8.18]{CLP16}
are a case in point; these allow both over- and underdispersion.

\subsection{FIC for linear mixed effects models}
\label{section:fic-lme}

\def\year{{\rm year}}

Models with random effects, often called mixed effect models,
are widely used in ecological applications. In
\citet*{CunenWalloeHjort19a} FIC formulae
have been developed for the class of {\it linear}
mixed effect models (often abbreviated LME models).
Here we will give a brief description
of that approach, which also serves as a special case
of the general FIC approach for a fixed wide model framework,
see \eqref{eq:fictwo}. Generalisations to classes
of nonlinear mixed effect models, and also to
heteroscedastic situations where variance parameters
depend on covariates, can be foreseen, following
similar chains of arguments but involving more
elaborations.

Suppose we have $n$ observations of $y_i$, a vector of length $m_i$.
The $m_i$ datapoints within each $y_i$ vector are assumed to be
dependent, and will often correspond to data collected in the same
space or time. Here we will refer to these data as
belonging to the same \textit{group}. Each $y_i$ vector
is associated with a regressor matrix $X_i$ of dimension $m_i\times p$
for the fixed effects, and a design matrix $Z_i$ of dimension
$m_i\times k$ for the random effects. The linear mixed effects
model takes the form
\beqn
y_i=X_i\beta+Z_ib_i+\eps_i
   \quad {\rm for\ }i=1,\ldots,n,
\eeqn
with the $b_i\sim\N_k(0,D)$ independent of the
errors $\eps_i\sim\N_{m_i}(0,\sigma^2I_{m_i})$.
The model may also be represented as
\beq
\label{eq:lme1}
Y_i\sim\N_{m_i}(X_i\beta,\sigma^2(I_{m_i}+Z_iDZ_i^\tr)),
\eeq
and its parameters are $\theta=(\beta,\sigma,D)$. Note that
the ordinary linear regression model is a special case,
corresponding to $D=0$.
The log-likelihood contribution for this group
of the data may be written
\beqn
\ell_i(\theta)
   =-m_i\log\sigma-\half\log|I_{m_i}+Z_iDZ_i^\tr|
   -\half(1/\sigma^2)(y_i-X_i\beta)^\tr(I_{m_i}+Z_iDZ_i^\tr)^{-1}(y_i-X_i\beta).
\eeqn
The combined log-likelihood $\sumin\ell_i(\theta)$
leads to maximum likelihood estimators
% $\hatt\theta_\wide=(\hatt\beta_\wide,\hatt\sigma_\wide,\hatt D_\wide)$
and hence also to
$\hatt\mu_\wide=\mu(\hatt\beta_\wide,\hatt\sigma_\wide,\hatt D_\wide)$
for any focus parameter $\mu=\mu(\beta,\sigma,D)$ of interest.

In applied situations we will spend efforts and
call on biological knowledge to construct a well-motivated
wide model, of the form \eqref{eq:lme1}.
This wide  model will typically be based on our
knowledge of the system under study and, crucially, on how
the data were collected. Quite often the resulting model
could become \textit{big}, in the sense that it includes
a large number $p$ of fixed effects
and also a large number $k$ of random effects.
Assume, as we do throughout this paper, that our primary interest
lies in the precise estimation of some focus parameter
$\mu$, which could be a function of the fixed
effect coefficients $\beta$, and/or the variance components
$(\sigma, D)$. For such a $\mu=\mu(\beta,\sigma,D)$,
can we find another model which offers more precise
estimates of $\mu$ than 
$\hatt\mu_\wide=\mu(\hatt\beta_\wide,\hatt\sigma_\wide,\hatt D_\wide)$
implied by the wide model?

FIC answers the question above; we can search among a set
of candidate models for one giving more precise estimates of $\mu$.
In the simplest setting, the candidate model
is defined with respect to the same $n$ groups
as in the wide model in \eqref{eq:lme1}, and we write
\beqn
%% \label{eq:lmeM}
y_i \sim \N_{m_i}\bigl(X_{M,i}\beta_{M},
   \sigma_{M}^2(I+Z_{M,i}D_{M} Z_{M,i}^\tr)\bigr).
\eeqn
This model has design matrices, $X_{M,i}$ and $Z_{M,i}$,
potentially different from those of the wide model,
and hence also a different set of parameters, say
$\theta_M=(\beta_M,\sigma_M,D_M)$.
Often, but not necessarily, the candidate model will
involve subsets of the covariates (i.e.~columns)
included in $X_i$ and $Z_i$, respectively. Let the
covariate matrix $X_{M,i}$ have dimension
$m_i\times p_M$, and $Z_{M,i}$ being $m_i\times k_M$.
The focus parameter must then be represented
properly inside the candidate model,
as $\mu_M=\mu_M(\beta_M,\sigma_M,D_M)$,
leading to the estimate
$\hatt\mu_M=\mu_M(\hatt\beta_M,\hatt\sigma_M,\hatt D_M)$.

In order to work out FIC formulae, we first need to study
the joint large-sample behaviour of the estimator
from the wide model $\hatt \mu_\wide$ and the estimator from the
candidate model $\hatt \mu_{M}$. 
This is as with eq.~(\ref{eq:approx3}) in 
Section \ref{subsection:fic-ii}, but the current
framework is more complicated and needs further efforts.
Such work is carried out in \citet*{CunenWalloeHjort19a}, 
and lead to 
\beqn
%% \label{eq:approx4}
\begin{pmatrix}
\rootn(\hatt\mu_\wide-\mu_\true) \\
\rootn(\hatt\mu_M-\mu_{0,M,n})
\end{pmatrix}
   \approx_d\N_2(0,\Sigma_{M,n}),
\eeqn
with all quantities defined analogously
to what is presented in Section \ref{subsection:fic-ii}.
These include matrices $J_n$, $J_{M,n}$, $K_{M,n}$, $C_{M,n}$
and gradient vectors $c$ and $c_{M,n}$,
defined similarly to those in Section \ref{subsection:fic-ii}, 
%% {\color{blue} in the appendix},
but here involving more complicated details
than for the plainer regression models worked with there.

This work then yields the same type of FIC formulae 
as for eq.~(\ref{eq:fictwo}), 
but with other recipes and formulae for the required
estimators for the quantities mentioned. 
% $c$, $c_M$, $J_n$, $J_{M,n}$, $K_{M,n}$, $C_{M,n}$. 
% Here we need to choose a good strategy in order to estimate all
% the necessary quantities. 
% This concerns in particular the matrices
% $J_n$, $J_{M,n}$, $K_{M,n}$, $C_{M,n}$ (see \eqref{eq:fictwo}). 
Regarding estimators for the matrices involved, 
we have three general possibilities:
(i) working out explicit formulae and
plug in the necessary parameter estimates;
(ii) computing the matrices numerically,
involving certain numerical integration details;
%% (this is what Jackson package does?);
(iii) via bootstrapping from the estimated wide model.
In \citet*{CunenWalloeHjort19a} the first option is pursued,
involving lengthy derivations of log-density derivatives
and their means, variances, covariances, computed
under the wide model. The resulting formulae are
too long for this review, but are fast to compute.
% Note that the derivations rely working with the
% definitions of the $J_n$,
% $J_M$, $K_M$ and $C_M$ as given in Section 2.2, and where all the
% expectation and variances are taken with respect to the wide model.
Options (ii) and (iii) have yet to be fully investigated,
but will likely be fruitful when extending this FIC approach
to the wider class of generalised linear mixed models
(the so-called GLMMs).
%% GLMMs.

The approach described here will be illustrated in
Section \ref{section:whales},
but we first offer some comments of a more general nature.
Readers familiar with linear mixed effects models
will be aware that there are two different estimation
schemes for models of this class, full maximum
likelihood and so-called REML estimators,
for restricted or residual maximum likelihood.
The REML method takes the estimation of the fixed effects
of the model into account when producing estimators
of the variance parameters.
For the computation of FIC scores the user might employ
either maximum likelihood or residual maximum likelihood
estimates, since these are large-sample equivalent;
see for instance \citet[Ch.~3]{Demidenko13}.
As with the general FIC formulae (\ref{eq:fictwo})
there are two versions, the approximately unbiased
estimates of risks and the adjusted ones.
In \citet*{CunenWalloeHjort19a} it is argued
that the unbiased version
\beq
\label{eq:ficforlme}
\fic_M^u=\hatt c_M^\tr\hatt J_M^{-1}\hatt K_M
   \hatt J_M^{-1}\hatt c_M/n
   +\hatt b_M^2-\hatt\kappa_M^2/n
\eeq
tends to work best for linear mixed effects models.
The benefit of this version is that good candidate models
with small biases earn more, compared to the wide model.
% The drawback is that it can lead
% to negative FIC scores (which we then can truncate to zero).
%%
Investigations show that the FIC formulae of (\ref{eq:ficforlme})
work well, in the sense that they accurately estimate the
risk associated with the use of the different candidate models.
The FIC formulae are based on large-sample arguments,
which for the case of the linear mixed effects models
involves approximations to normality when the number $n$
of groups increases to infinity. These normal approximations
work well as long as the full sample size $\sumin m_i$ grows,
particularly for functions of the linear mean parameters.
More care is sometimes required when it comes to
applications involving non-linear functions of
both mean and variance parameters, as with
estimates of probabilities $\mu=\Pr(Y\ge y_0\midd x_0,z_0)$.

\section{Application: The slimming of minke whales}
\label{section:whales}

%% \label{subsection:whales}

%\begin{figure}[h]
%%% \label{figure:whales1}
%\begin{center}
%\includegraphics[scale=0.4]{frontiers_whalesfig1}
%\end{center}
%\caption{Estimates of the yearly decline in fat-weight
%focus parameter, for the Antarctic minke whale population
%(vertical axis), along with root-FIC scores (horizontal axis),
%for the wide model $M_0$, marked in blue, and five
%additional candidate models $M_1,\ldots,M_5$.
%The scale is in kilograms of fat.}
%\label{figure:whales1}
%\end{figure}

Our second application story concerns the potential change
in body condition of Antarctic minke whales over a period of 18 years.
For a more thorough investigation consult
\citet*{CunenWalloeHjort19b}. Questions treated there
have been discussed in the Scientific committee
of the International Whaling Commission (IWC)
for a number of years, and a full consensus has
not been reached. In the context of this review, therefore,
the analysis below should be taken as an illustration,
and not necessarily the last word on the topic
of the decline in energy storage or body condition
for the minke whales.

Using data from the Japanese Whale Research Program
under Special Permit in the Antarctic (the so-called JARPA-1)
we have studied the evolution of fat weight in Antarctic
minke whales caught in 18 consecutive years, from 1988 and 2005.
The main biological interest lies in whether or not
the whales experienced a decline in body condition during
the study period, and the dissected fat weight (in tonnes or kg)
is taken to be a proxy for this body condition.
Thus, there is a clear focus parameter in this
application: the yearly decline in fat weight (which we will
parametrise in a suitable fashion in the following).

The whales caught in each year are unevenly sampled
with respect to a number of covariates, for instance sex,
body length, age, and longitudinal region in the Antarctic ocean.
Since all these covariates may  influence body
condition we need to include them in a model aiming
at estimating the potential yearly decline in the response.
Based on lengthy and detailed discussions in the
Scientific Committee of the IWC,
we have chosen a wide model within the class of linear
mixed effect models, see Section \ref{section:fic-lme}.
In \citet*{CunenWalloeHjort19b}
we have used considerable efforts to motivate
the choice of covariates, interactions, and random effect
terms in the wide model, but these arguments are outside
the scope of the present article.
In R-package-type notation, the wide model
can be given as
\beqn
\fatweight \sim
&&\!\!\yeartt + \yeartt^\twott + \bodylength + \sex + \diatom
   + \datett +  \datett^\twott + \age \\
&& +\,  \sex*\diatom + \diatom*\datett +  \diatom*\datett^\twott
   + \bodylength*\sex \\
&& +\,  \bodylength*\datett +  \bodylength*\datett^\twott
+ \sex*\datett + \sex*\datett^\twott \\
&& +\,   \bodylength*\sex*\datett + \bodylength*\sex*\datett^\twott + \age*\sex  \\
&& +\, \age*\datett + \age*\datett^\twott + \age*\sex*\datett + \age*\sex*\datett^\twott  \\
&& +\, \yeartt*\sex  + \yeartt^\twott*\sex + \region + \yeartt*\region + \yeartt^\twott*\region  \\
&& +\, \sex*\region +  \diatom*\region + \region*\datett \\
&& +\,  \region*\datett^\twott  + (1 + \datett + \datett^\twott\midd\yeartt).
\eeqn
The ${\region}$ covariate reflects three different geographical
regions, associated with three regression coefficients
summing to zero.

The model defined above has $p=40$ fixed effect coefficients.
The notation $(1 + \datett + \datett^\twott\midd \yeartt)$ specifies
the random effect structure; the groups are defined by
a categorical version of the year variable (so $n=18$),
and the $Z_i$ matrix has $k=3$ columns (a column of ones for
the intercept, date, and date squared).
According to prior biological knowledge, $\datett$ is assumed to be
one of the most important effects governing the fat weight.
The variable refers to the day of the season when each whale
was caught, and since the whales are in the Antarctic
to gain weight the coefficient related to date is expected
to be large and positive. Also, the effect of date
is expected to be different from year to year,
possibly due to fluctuations in krill production.
Hence, a random effect on $\datett$ is included. We thus
have a total of $40+1+6=47$ parameters to estimate.
The total number of observations, i.e.~$\sumin n_i$, was 683.

As mentioned above the main interest, for discussions
at several IWC meetings, has been the yearly decline
in the $\fatweight$ outcome variable. Since we have
a quadratic year term in our wide model, with that part
taking the form $\beta_\year x+\beta_{\year2}x^2$ for year $x$,
a natural definition of the yearly decline is
$\mu = \beta_{\year}+2\beta_{\year2} x_0$,
with $x_0$ the mean year in the dataset.
The focus parameter corresponds to the derivative of
the mean response, with respect to year,
and evaluated in this mean year time point.
For candidate models with only a linear effect of year
the parameter simplifies to $\beta_{\year}$ only.
Furthermore, for those submodels where there is no year effect
included, we have $\beta_\year=0$, a parameter value
which then is estimated with zero variance but
with potentially big bias.
For this example, we have limited ourselves
to investigating five candidate models only, in addition
to using the wide model itself; see Table \ref{table:table1}.

We do not actually expect the mean level of decline
in energy storage to be either exactly linear
or exactly quadratic over 18 years, but take
this level of approximation to be adequate for the
purpose, since the decline over time curve is
not far from zero; also, our focus parameter
is identical to the overall slope, the mean curve
evaluated at the end point minus its value at the
start point, divided by the length of time.
% and $\beta_{\year2}$ is the coefficient corresponding to the
% second order year term.

All the candidate models have a smaller number of fixed effects
than the wide model. Note that the first candidate model $M_1$
has a more complex random effect structure than the wide model itself
(with $k=6$ giving a total of 21 random effect parameters).
This choice also demonstrates that there is nothing
in the formulae hindering us from having candidate models
with more random effects (or also more fixed effects)
than the wide model. When it comes to interpreting the results,
it is usually more natural to choose the wide model to be
the largest possible plausible model, however.
The models $M_2$ and $M_3$ are very simple (with few fixed effects),
and differ only in the their random effects. Model $M_4$ includes only
the linear year effect in addition to a single random effect in the
intercept. The last model, $M_5$, is the model without any year effect,
so $\mu_{M_5}=0$. With the present focus parameter, the FIC score
of such  a model will have zero variance and a bias which
only depends on the estimated focus parameter
in the wide model, and its estimated variance, so
$\fic^u_{M_5} = (0-\hatt\mu_\wide)^2-\hatt\kappa_\wide^2/n$,
for the relevant $\kappa_\wide^2/n$ approximation
to the variance of $\hatt\mu_\wide$. %; cf.~equation (\ref{eq:twofic}).
Thus, further specification of $M_5$ is unnecessary;
it includes all possible LME models without any year effect.
As the candidate models worked with are not close enough to each other to warrant
the use of the local neighbourhoods framework, we use the ‘fixed wide model’ approach.

\begin{table}[h]
\small
\begin{center}
\begin{tabular}{ccccc}
% \toprule
                & description & $p$ & $k$ & $d$ \\ \midrule
    $M_0$ & wide model & 40 & 3 & 47 \\
    $M_1$ & less interactions, quadratic year effect & 9 &  6 & 31 \\
    $M_2$  &  very simple, linear year effect & 5 & 2  & 9\\
    $M_3$ &  very simple, linear year effect  & 5 & 1 & 7 \\
    $M_4$ &  only linear year effect  & 2 & 1 & 4\\
    $M_5$ &  like the wide, but without year effect  & 32 & 3 & 39 \\
% \bottomrule
\end{tabular}
\end{center}
\caption{Brief descriptions of the wide model and the five
additional candidate models, with the number of fixed effects,
the number of random effects, and the total number of parameters
to be estimated, for each model.}
\label{table:table1}
\end{table}

\begin{figure}[h]
%% \label{figure:whales1}
\begin{center}
\begin{tabular}{cc}
(a) & (b) \\
\includegraphics[scale=0.31]{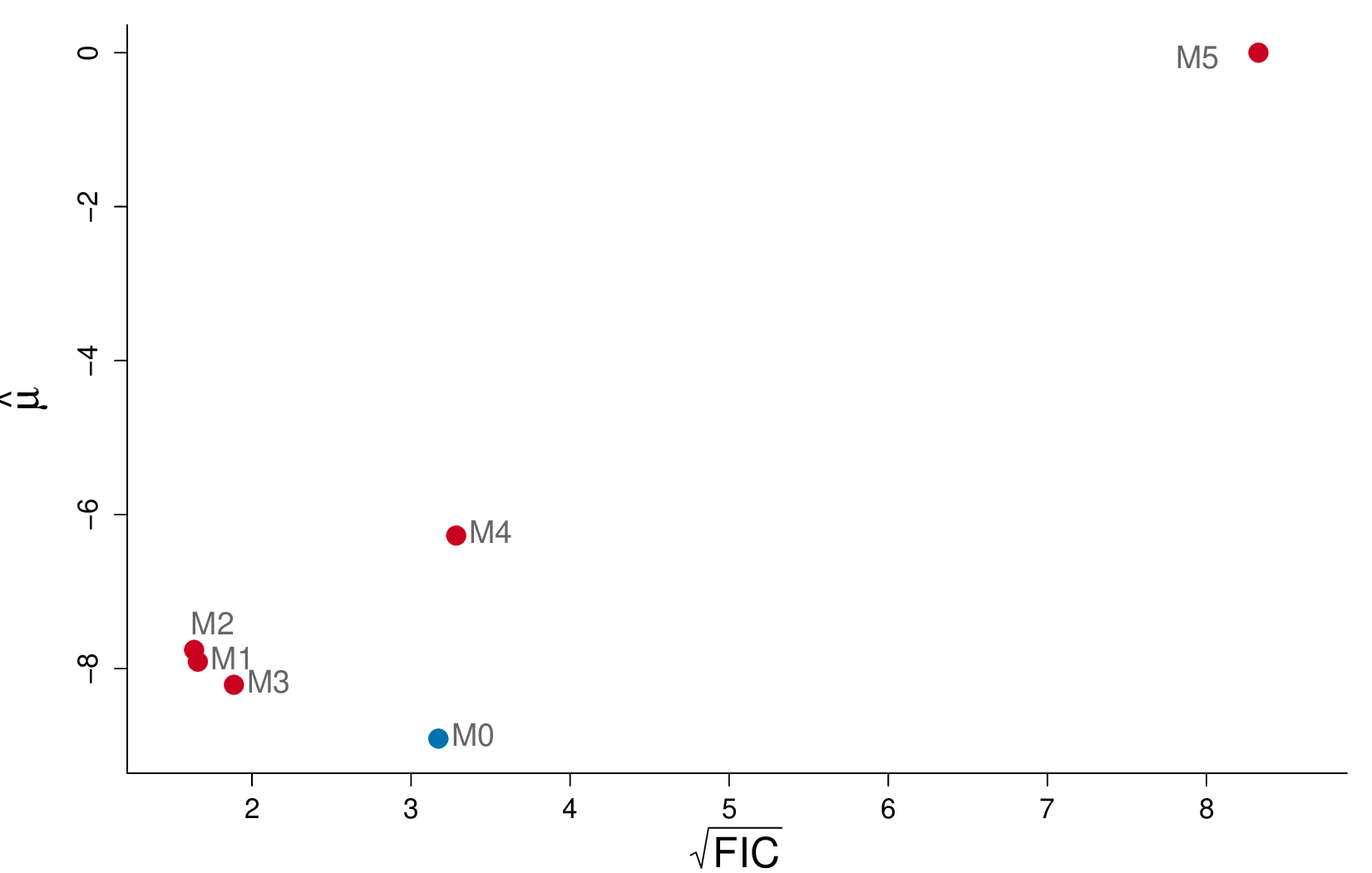} &
\includegraphics[scale=0.31]{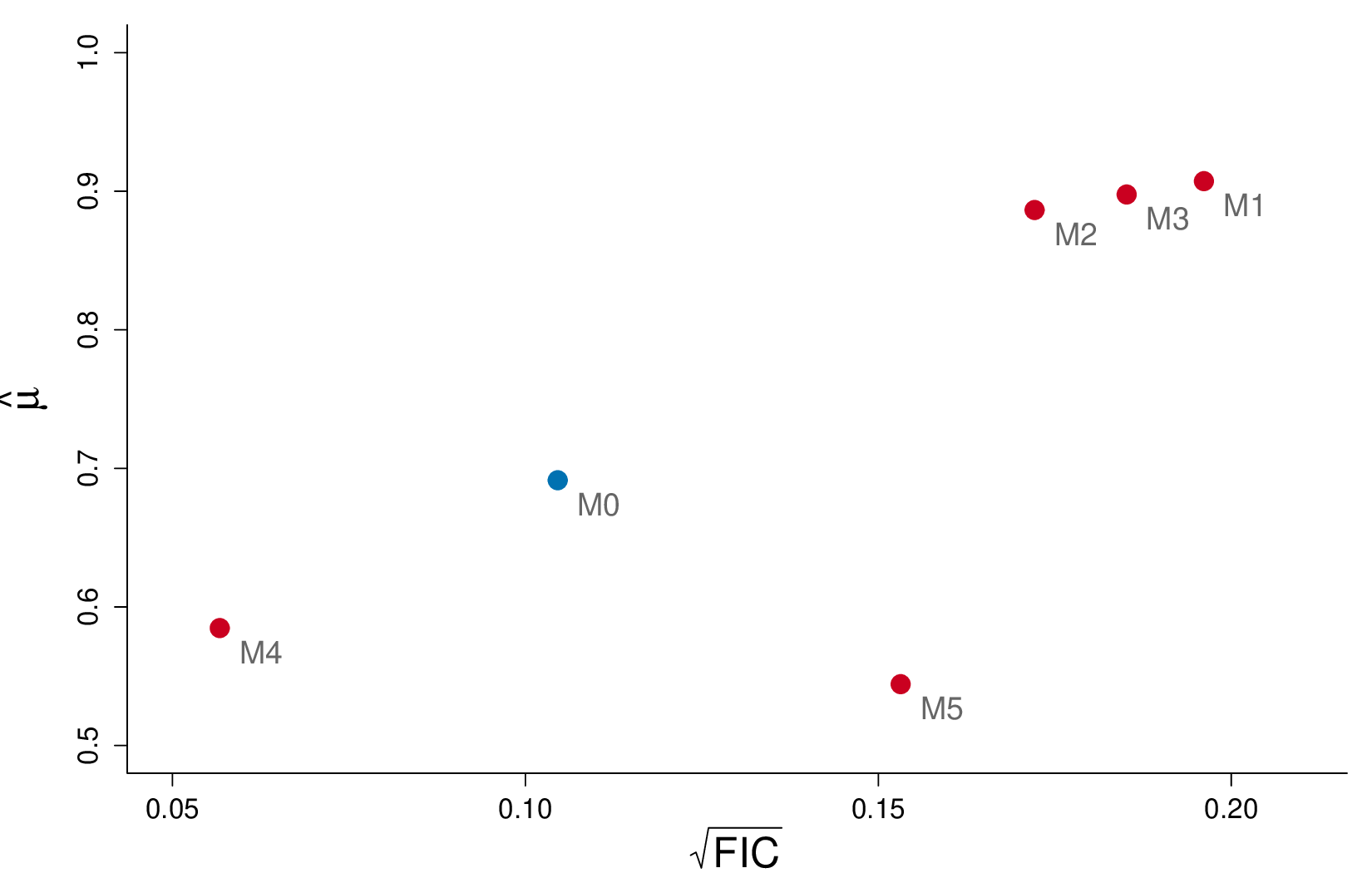}
\end{tabular}
\end{center}
\caption{(a) Estimates of the yearly decline in fat-weight
focus parameter, for the Antarctic minke whale population
(vertical axis), along with root-FIC scores (horizontal axis),
for the wide model $M_0$, marked in blue, and five
additional candidate models $M_1,\ldots,M_5$.
The scale is in kilograms of fat.
(b) Root-FIC scores and estimates of the probability of observing
a whale with more than 1.5 tonnes of fat for the wide model
(marked in blue) and the five candidate models.
}
\label{figure:whales1}
\end{figure}

%\begin{figure}[h]
%\begin{center}
%\includegraphics[scale=0.4]{frontiers_whalesfig2}
%\end{center}
%\caption{Root-FIC scores and estimates of the probability of observing
%a whale with more than 1.5 tonnes of fat for the wide model
%(marked in blue) and the five candidate models.}
%\label{figure:whales2}
%\end{figure}

After carefully constructing our wide model,
and checked that it passes various diagnostic tests,
we can proceed to model selection with the FIC.
The results  are given in the form of a FIC-plot
in Figure \ref{figure:whales1}. We see that $M_2$
gets the lowest FIC score, with $\hatt\mu=-7.76$.
The models $M_1$ and $M_3$ are close to the winning
one, both in terms of their FIC scores and their
estimates of the focus parameter.
Model $M_5$, without the year effect had a considerably
larger FIC score than any of the other models
(which can be seen as an implicit test for the
the null hypothesis of there being no year effect).
From the plot we can conclude that our best estimate
of the focus parameter is around $-8$ kilograms,
or 80 kg loss of fat over a decade. Furthermore,
since the root-FIC values are about $1.50$,
confidence intervals associated with these best
point estimates will clearly fall to the left of zero.
A natural interpretation of the FIC plot is therefore
that the body condition decline, for the Antarctic
minke whales, has been {\it negative and significant}
over the study period.

To demonstrate the versatility of our approach, we have investigated
the same six models with respect to another focus parameter,
the probability of observing a whale with more than
a certain amount of fat, say 1.5 tonnes (1500 kg),
given some covariate values:
$\mu_2=\Pr(Y \ge 1.5 \midd x_0,z_0)$. Here we chose to look
at a 20 year old male whale, caught in 1991 in the eastern
region, of approximately mean length (8 metres),
and which is caught towards the end of the season.
Over the full dataset, the average fat weight
of a whale is close to 1.5 tonnes.
The FIC scores and estimates are given in Figure \ref{figure:whales1}.
We observe that the models give widely different estimates,
ranging from around 0.50 to 0.90, and that the ranking
of the models is very different from the ranking when
the focus was the yearly decline in fat weight.
The smallest model $M_4$ is considered the best for
estimating the probability of observing a `medium fat'
whale. Here, we see the typical bias-variance trade-off
at work: using $M_4$ clearly gives an estimate
with some bias compared to the wide model
(estimate of 0.60 instead of around 0.70),
but the bias is compensated for by a
strong decrease in variance.

\section{Discussion}
\label{section:discussion}

Our article has motivated, exhibited, developed, and extended 
the machinery of Focused Information Criteria for model selection
and model ranking, with a few illustrations for
ecological data. Here we offer some general remarks.

{\it 1. The role of the wide model.}
The FIC idea is to examine how different candidate models
work regarding what they actually deliver, in terms
of point estimates for the most crucial parameters
of interest. This examination involves approximations
to and estimates of the risks, which for the
usual squared error loss function means mean squared
error. Quantifying the implied variances and biases
relies on the notion of a clearly defined
(though unknown) data generating mechanism.
This is one of the roles of our {\it wide model}.
In the local asymptotics framework of
Section \ref{subsection:fic-i} this is the
full model $f(y_i\midd x_i,\theta,\gamma)$
of (\ref{eq:ftruedelta2}), with $p+q$ parameters;
in the alternative framework of
Section \ref{subsection:fic-ii} it is what
we term the fixed wide model.
Such a wide model needs to be well argued,
as being sufficiently rich to encompass the anticipated
salient features of the phenomena studied.
Since quantification and consequent estimation
of variances and biases rest on the wide model
being adequate it ought also to be given
a goodness-of-fit verification, involving
diagnostic checks etc.

%% {\color{blue}
One might inquire how sensitive the FIC scores are to the choice 
of the wide model.  In connection with the application 
described in Section \ref{section:whales} we have 
conducted some sensitivity checks and found that moderate 
changes to the wide model had little effect on the ranking 
of the different candidate models. Also, for the wide models 
we have investigated, the estimate of the focus parameter 
in the selected models was reasonably stable. More radical 
changes to the wide model should be expected to have greater 
effect, but we have not fully investigated this issue. 
Fully guarding against all misspecification of the 
wide model is unattainable, but extending our approach 
to even wider and more flexible wide models may lead to some improvements.

{\it 2. When should you use FIC?}
Practitioners may be interested in model selection for different,
overlapping reasons. On one hand the goal might be to select
the candidate model which in a relevant sense
is the closest to the true data generating mechanism.
Criteria based    on model fit and some penalisation
for complexity aim at this goal,
for instance the well-known AIC and BIC;
see \citet{ClaeskensHjort08a} for a general discussion.
On the other hand, practitioners often seek a small model
offering precise estimates of the quantities
they are interested in.
It is important to keep in mind that FIC specifically aims
at the second goal, and is not necessarily suitable
for the first goal. FIC offers a principled way to
{\it simplify} a large, realistic model which the
user assumes to hold (i.e.~to be realistically
and adequately close to the complicated truth).
%% (or at least close to the truth).
The goal of the simplification is to obtain more precise estimates
of quantities of interest, say $\hatt\mu$ for an underlying
focus parameter $\mu$. This also includes producing predictions
for not yet seen outcomes of random variables, like
the abundance of a certain species over the coming twenty years.
Here simplification must be understood in a wide
sense, as the candidate models do not necessarily
need to be nested within the wide model, as we have seen.
The two different motivations for model selection alluded
to above partly relate to the two goals for statistical modelling:
to explain or to predict, i.e.~the `two cultures of statistics',
see \citet{Breiman01, Shmueli10}. For yet further perspectives
on model selection with focused views, coupled with 
model structure adequacy analysis, see \citet{TaperStaples08}. 

%% {\color{blue}
Once a practitioner has decided to use FIC, she then has to 
make a choice  between the two FIC frameworks we have discussed, 
using local asymptotics or a fixed wide model. As a tentative 
guiding rule we advocate turning to the `fixed wide model' 
setup if the set of candidate models are seen as not being 
in a reasonable vicinity of each other. Also,
we have seen that this  framework allows 
candidate models of a different sort from the 
wide model; in particular, a candidate model does
not have to be a clear submodel of the wide model. 
As stated before, the two frameworks aim at the same quantities, and 
the choice may thus also be guided by convenience. Note also that in 
many situations the two frameworks may give similar results. 
%For given classes of models and estimands the two
%frameworks can of course be compared. 
For the
special case of linear regression models
%,say $y_i\sim\N(x_i^\tr\beta+z_i^\tr\gamma,\sigma^2)$,
with focus parameters being linear functions 
of the coefficients, the formulae turn out to
be identical. Also, for the classical generalised
linear models, including logistic and Poisson
regressions, the formulae yield highly correlated
scores, as long as the focus parameters under study
are functions of such linear combinations
$x_0^\tr\beta+z_0^\tr\gamma$. For more complicated
focus parameters, like probabilities for crossing
thresholds, the answers are not necessarily close,
and will depend on both the sample size and
the degree to which the candidate models are
not close.

{\it 3. Model averaging.}
Model averaging is sometimes used as an alternative to model selection
to avoid the perhaps brutal throwing away of all but one model. With
model averaging one computes the estimate of the focus quantity in all
of the models separately and then forms a weighted average which is used
as the final `model averaged' estimate of the focus.
See for example the overview paper about model averaging in ecology by
\citet{Dormannetal18}.
Averaging estimates has as the advantage that all models are used. The
flexibility of choosing the weights allows to give a larger weight to
the estimate of a model that one prefers most. Weights could be set in a
deterministic way, such as giving equal weights to all estimates, or
could be data-driven. It makes sense to use values of information
criteria to set the weights. Especially AIC has been popular, see
\citet{BurnhamAnderson02} for examples of the use of `Akaike weights'.
Also FIC could be used to form weights that are proportional to
$\exp(-\lambda\,\fic_M/\fic_\wide)$ for a user-chosen value of $\lambda$.
One could also try to set the weights such that the mean squared error
of the weighted estimator is as small as possible
\citep{Liangetal11}.
Such theoretically optimal weights need to be estimated for practical
use, which induces again estimation variability, and might lead to a
more variable weighted estimator as when simple equal weights would have
been used \citep{Claeskensetal16}.

Model averaging with data-driven weights has consequences for
inference similar to the post-selection inference
(see below). Indeed, model selection may be seen as
a form of model averaging, with all but one of the weights
equal to zero and the remaining weight equal to one.
Correct frequentist inference for model
averaged estimators needs to take the correlations
between the separate estimators into account,
as well as the randomness of the weights in case of
data-driven weights.

{\it 4. Post-selection issues.}
Model selection by the use of an information criterion
(such as FIC, or AIC) comes with several advantages as
compared to contrasting models two by two via hypothesis
testing. With model selection there is no need
to single out one model that would be placed in a null
hypothesis. All models are treated equally. Multiple testing
issues do not occur because no testing takes place.
The set of models that is searched over can be large.
The ease of calculating such information criteria makes
it fast and allows to include many models in the search.
However, there is a price to pay when one puts the next
step to perform inference using the selected model.
Simply ignoring that a model is arrived at via a selection
procedure results in p-values that are too small and
confidence intervals that are too narrow.

With a replicated study resulting in a dataset similar to but
independent of the current one, it might happen that a different model
gets selected, all the rest left unchanged. This illustrates that
variability is involved in the process of model selection.
One way to address such variability is via model averaging;
see e.g.~\citet{HjortClaeskens03}, \citet[Ch.~7]{ClaeskensHjort08a},
\citet{Efron14}. \citet{Berketal13} develop an approach for the
construction of confidence intervals for parameters in a linear
regression model that uses a selected model. Their approach is
conservative, in the sense that the intervals tend to be wide and
sometimes have a coverage that is quite a bit larger than the nominal
value. Other approaches to take the uncertainty induced by the selection
procedure into account is via selective inference leading to so-called
`valid' inference. See, for example,
\citet{Tibshiranietal16, Tibshiranietal18}. By
using information about the specifics of the selection method such
inference methods result in narrower confidence intervals as compared
to the \citet{Berketal13} method. The effect of increasing the number of
models results in getting larger confidence intervals, see
\citet{CharkhiClaeskens18}. Valid inference after selection is currently
investigated for several model selection methods. It is to be expected
that more results will become available in the future that guarantee
that working with a selected model happens in a honest way that takes
all variability into account.

It is well known that estimators computed under a given model
become approximately normal, under mild regularity
conditions. It is however clear from the brief discussion
above that post-selection and more general model-average
estimators have more complicated distributions,
as they often are non-linear mixtures of approximately
normal distributions, with different biases, variances,
and correlations. Clear descriptions of large-sample behaviour,
for even complex model-selection and model-average schemes,
can be given inside the local asymptotics $O(1/\rootn)$
framework of Section \ref{subsection:fic-i},
as shown in \citet{HjortClaeskens03},
\citet[Ch.~7]{ClaeskensHjort08a}, with further
generalisations in \citet{Hjort14}. Inside the general
framework of (\ref{eq:ftruedelta2}), with estimators
$\hatt\mu_M$ as in (\ref{eq:muhatM}), consider
the combined or post-selection estimator
\beqn
\hatt\mu^*=\sum_M \hatt w(M)\hatt\mu_M,
\eeqn
with data-driven weights $\hatt w(M)$ summing to one.
If these are weights take the form $w(M\midd D_n)$,
with $D_n=\rootn(\hatt\gamma_\wide-\gamma_0)$ as
in (\ref{eq:hereisDn}), there is a very clear limit distribution,
\beq
\label{eq:masterII}
\rootn(\hatt\mu^*-\mu_\true)
\arr_d \Lambda_0+\omega^\tr\{\delta-\hatt\delta(D)\},
\quad {\rm where} \quad
\hatt\delta(D)=\sum_M w(M\midd D)G_MD.
\eeq
This extends the master theorem
result (\ref{eq:master1}), to allow even for very complicated
post-selection and model averaging schemes.
The $q\times q$ matrices $G_M$ in this orthogonal
decomposition are as in (\ref{eq:hereisGM}.
The result remains true also for schemes based on weights
involving AIC or FIC weights, as the appropriate
weights can be shown to be close enough to the
relevant $w(M\midd D_n)$.
These limiting distributions can be simulated,
at any position in the $\delta$ domain. Yet further
efforts are required to turn such into valid
post-selection or post-averaging confidence intervals,
however; see \citet[Ch.~7]{ClaeskensHjort08a}
for one particular general (conservative) recipe,
and for further discussion of these issues.

{\it 5. Performance.}
It is beyond the scope of this article to go into
the relevant aspects of statistical performance
of the FIC methods. One may indeed study both
the accuracy of the final post-selection or
post-averaged estimator, say for the $\hatt\mu^*$
above, and the probabilities for selecting the 
%% {\color{blue} truly}
best models. Such questions are to some extent
discussed in \citet{HjortClaeskens03} and
\citet[Ch.~7]{ClaeskensHjort08a}; broadly speaking,
the FIC outperforms the AIC in large parts of
the parameter space, but not uniformly. There
are also several advantages with FIC, when
compared with the BIC, regarding precision of
the finally evaluated estimators. Notably,
all of these questions can be studied accurately
in the limit experiment alluded to above,
where all limit distributions can be given
in terms of the orthogonal decomposition
$\Lambda_0+\omega^\tr\{\delta-\hatt\delta(D)\}$
of (\ref{eq:masterII}).

{\it 6. FIC for high-dimensional data.}
When models contain a large number of parameters, perhaps 
even larger than the sample size, maximum
likelihood estimation might no longer be appropriate. 
The use of regularised estimators, such as ridge regression,
lasso, scad, etc.~requires adjustment to the FIC formulae. 
Even when the regularisation takes automatic care of selection, 
\citet{Claeskens2012-PenFIC} showed that selection via FIC is
advantageous to get better estimators of the focus. 
\citet{Pircalabelu-etal2016} used FIC for high-dimensional 
graphical models. For models with a diverging number of 
parameters FIC formulae using a so-called desparsified 
estimator have been obtained by \citet{GueuningClaeskens2018}.
FIC may also be used to select tuning parameters for ridge
regression. The focused ridge procedure of \citet{HelltonHjort2018}
is applicable to both the low and high-dimensional case and has
been illustrated in linear and logistic regression models. 

{\it 7. Extensions to yet other models.}
The methods exposited in Section \ref{subsection:fic-ii},
%% and \ref{section:fic-lme},
yielding FIC machinery under a fixed wide model,
can be extended to other important classes of models.
The essential assumptions are those related to smooth
log-likelihood functions and approximate
normality for maximum likelihood estimators
for the candidate models. Sometimes developing
such FIC methods would take considerable
extra efforts, though, as exemplified by
our treatment in Section \ref{section:fic-lme}
of linear mixed effects models.
In particular, the methodology extends to
models with dependence, as for time series and
Markov chains with covariates, see \citet{Haug19}.
This involves certain lengthier efforts
regarding deriving expressions and estimation
methods for the $K_{M,n}$ and $C_{M,n}$
matrices of (\ref{eq:approx2})--(\ref{eq:approx3}).
Analogous FIC methods for time series are
shown at work in \citet*{HermansenHjortKjesbu16}
for certain applications in fisheries sciences.
Similar remarks also apply to the advanced Ornstein--Uhlenbeck
process models used in \citet*{ReitanSchwederHendericks12}
for modelling complex layered long-term evolutionary data.
Specifically, these authors studied cell size evolution
over 57 million years, and entertained 710 candidate
models of this sort. An extension of our paper's
FIC methods to their process models is possible
and would lead to additional insights in their data.

A challenge of a different sort is to develop
FIC methods also when the models used are too
complicated for log-likelihood analyses, but
where different estimation methods may be used.
A case in point are models used in \citet{DennisTaper94},
for dynamically evolving times series models
of the form $y_{t+1}=y_t+a+b\exp(y_t)+\sigma z_t$,
met in density dependence analyses for ecology.
These models do not have stationary distributions
and special estimation methods are needed to
analyse the candidate models.

% \appendix

\section{Appendix: Derivations and technical details}
\label{section:details}

%% \section{Appendix: Derivations, proofs, and the bird species data}

%% subsection: fic-i         which does O(1/\rootn) 
%% subsection: fic-glm-local which does FIC for GLM for O(1/\rootn) 
%% subsection: 

Here we give some of the technical details and mathematical 
arguments, related to 
(i) FIC within the local neighbourhood framework 
for regression models, 
(ii) application of such methods and formulae for 
generalised linear models, and 
(iii) FIC for regression models using the fixed wide model framework
(see Sections \ref{subsection:fic-ii}--\ref{subsection:fic-iiii}). 
 
\subsection{FIC within a local asymptotic framework}
\label{subsection:fic-i}

In a local neighborhood framework one assumes that
regression data $(x_i,y_i)$ for $i=1,\ldots,n$ have true densities
\beq
\label{eq:ftruedelta2}
f_\true(y_i\midd x_i)=f(y_i\midd x_i,\theta_0,\gamma_0+\delta/\rootn),
\eeq
with $\theta$ of dimension $p$, and
$\gamma=\gamma_0+\delta/\rootn$ of dimension $q$.
The most simple model, the narrow model, has density
$f_\narr(y_i\midd x_i)=f(y_i\midd x_i,\theta_0,\gamma_0)$,
where $\gamma_0$ is known and $\theta_0$ is the unknown but true
value of this parameter.
For example, a narrow model might include only the 
intercept $\theta_0$ for the mean, setting all other regression 
coefficients that are present in a wide model equal to zero,
$\gamma_0=0$. The notation allows for more generality where, 
for example, scale parameters can be set to known
values under the narrow model. The wide model has $p+q$ 
parameters. Submodels of the wide model assume 
some of the components of $\gamma$ to be equal to the 
components of $\gamma_0$, while others are free to be
estimated. All models in the search procedure are in 
between the narrow and wide model.

We may now summarise basic results reached in
\citet{ClaeskensHjort03, ClaeskensHjort08a, HjortClaeskens03},
pertaining to estimation in all of these $2^q$ candidate models.
Let $\mu=\mu(\theta,\gamma)$ be a focus parameter,
and consider a candidate model $M$, identified
as the subset of $\{1,\ldots,q\}$ for which
the corresponding extra parameter $\gamma_j$
is inside the model, with $\gamma_j=0$ for $j\notin M$.
Maximum likelihood estimation inside this model $M$
leads to $(\hatt\theta_M,\hatt\gamma_M)$,
of dimension $p+|M|$, writing $|M|$ for the
number of elements of $M$. The ensuing estimate
for the focus parameter is
\beq
\label{eq:muhatM}
\hatt\mu_M=\mu(\hatt\theta_M,\hatt\gamma_S,\gamma_{0,M^c}),
\eeq
aiming for $\mu_\true=\mu(\theta_0,\gamma_0+\delta/\rootn)$.
In particular, maximum likelihood estimation in the wide model
leads to $\hatt\mu_\wide=\mu(\hatt\theta_\wide,\hatt\gamma_\wide)$.
Including all or many extra parameters in $M$
means low modelling bias but higher variance;
using only few means potentially bigger modelling bias
but smaller variance.

Large-sample theory for the full ensemble of these
estimators may now be worked out, via careful
refinements of traditional under-the-model methods,
along with a fair amount of algebraic efforts. Among
the chief results is the following, which needs
a bit of introduction to explain its ingredients.
First, writing $Y_i$ for the random variable
in question, consider
\beqn
%% \label{eq:fisher1}
J_n=n^{-1}\sumin \Var_0\,u(Y_i\midd x_i,\theta_0,\gamma_0)
   =\begin{pmatrix} J_{n,00} &J_{n,01} \\ J_{n,10} &J_{n,11} \end{pmatrix},
\eeqn
the Fisher information matrix, of size $(p+q)\times(p+q)$,
evaluated at the null point $(\theta_0,\gamma_0)$; here
% \beqn
$u(y_i\midd x_i,\theta,\gamma)
   =\dell\log f(y_i\midd x_i,\theta,\gamma)/\dell\eta$
% \eeqn
is the $p+q$-dimensional score vector, writing $\eta$
for the full model parameter vector $(\theta,\gamma)$.
The $J_n$ will converge to a well-defined positive definite
limit matrix $J_\wide$ under mild ergodic assumptions,
and with blocks $J_{00},J_{01},J_{10},J_{11}$.
Let next $Q_n=J_n^{11}$ be the $q\times q$
lower-right block of $J_n^{-1}$, along with its limit $Q$,
and define
\beq
\label{eq:omega}
\omega_n=J_{n,10}J_{n,00}^{-1}\dellone-\delltwo,
\eeq
a $q$ vector transformation of the partial derivatives
of $\mu(\theta,\gamma)$ with respect to the $\theta$
and $\gamma$ parameters, again evaluated at the null point.
Introduce independent limit variables
$\Lambda_0\sim\N(0,\tau_0^2)$ and $D\sim\N_q(\delta,Q)$,
with $\tau_0^2=(\dellone)^\tr J_{00}^{-1}\dellone$.
Here $D$ is the limit distribution version of
\beq
\label{eq:hereisDn}
D_n=\rootn(\hatt\gamma_\wide-\gamma_0).
\eeq
Finally we need to introduce the $q\times q$ matrices
\beq
\label{eq:hereisGM}
G_{M,n}=\pi_M^\tr Q_{M,n}\pi_M Q_n^{-1}
      =\pi_M^\tr (\pi_M Q_n^{-1}\pi_M)^{-1}\pi_M Q_n^{-1},
\eeq
where $\pi_M$ is the $|M|\times q$ projection matrix
taking $v=(v_1,\ldots,v_q)$ to the vector $v_M$
with only those $v_j$ for which $j\in M$.
These matrices become `fatter' with bigger subsets $M$,
and the trace of $G_{M,n}$ is simply $|M|$;
also, $G_{\emptyset,n}=0$ and $G_{\wide,n}=I_q$.
The limits $J$ and $Q$ of $J_n$ and $Q_n$ imply
corresponding limits $G_M$ of the $G_{n,M}$.

The master theorem for focus parameter estimators
in all these submodels says that
\beq
\label{eq:master1}
\rootn(\hatt\mu_M-\mu_\true)
   \arr_d\Lambda_M=\Lambda_0+\omega^\tr(G_MD-\delta).
\eeq
The convergence holds jointly, for the full ensemble
of $2^q$ candidate estimators of $\mu$. The
distribution of each $\Lambda_M$ is normal,
with means and variances depending on
the population quantities
$\tau_0$, $Q$, $\omega$, and the local
departure parameter $\delta$ associated with
$\delta/\rootn$ of (\ref{eq:ftruedelta2}).
Note that different foci $\mu$ have different
$\omega$ of (\ref{eq:omega}).

From these efforts follow clear expressions for the limiting
mse of all candidate estimators, as
\beqn
\mse(M)=\E\,\Lambda_M^2
   =\tau_0^2+\omega^\tr G_MQG_M^\tr\omega
   +\{\omega^\tr(G_M-I_q)\delta\}^2.
\eeqn
FIC formulae follow from this by estimating the required
population quantities. From the estimator
$\hatt J_n=-n^{-1}\dell^2\ell_n(\hatt\theta,\hatt\gamma)
   /\dell\eta\,\dell\eta^\tr$,
the normalised Hessian matrix associated with finding
the maximum likelihood estimators in the wide model,
follow estimates of its blocks and its inverse,
and hence $\hatt Q=(\hatt J_n^{-1})_{11}$.
Similarly $\hatt\omega$ can be put up by
taking derivatives of $\mu(\theta,\gamma)$ at the
wide model maximum likelihood position.
Note that $\E\,DD^\tr=\delta\delta^\tr+Q$,
so estimating $(c^\tr\delta)^2$ unbiasedly is achieved
by using $c^\tr(DD^\tr-Q)c=(c^\tr D)^2-c^\tr Qc$.
%% Truncating negative estimates of squared biases to zero,
All of this leads to the FIC formula for $\hatt\mu_M$, 
which uses an asymptotically unbiased estimator of the squared bias,  
\beq
\label{eq:ficone}
\begin{array}{rcl}
\fic_M^u
   &=&\displaystyle
n^{-1}\{\hatt\tau_0^2+\hatt\omega^\tr \hatt G_M\hatt Q\hatt G_M^\tr\hatt\omega
   +\hatt\omega^\tr(\hatt G_M-I_q)(D_nD_n^\tr-\hatt Q)
    (\hatt G_M-I_q)^\tr\hatt\omega\}.
\end{array}
\eeq
As explained in the general case of (\ref{eq:twofic}), 
a useful variant is to truncate any negative estimates
of squared biases to zero, leading to the adjusted
FIC, i.e.
\beqn
\fic_M=n^{-1}\{\hatt\tau_0^2+\hatt\omega^\tr \hatt G_M\hatt Q\hatt G_M^\tr\hatt\omega\}
   +\max\{n^{-1}\hatt\omega^\tr(\hatt G_M-I_q)(D_nD_n^\tr-\hatt Q)
    (\hatt G_M-I_q)^\tr\hatt\omega,0\}.
\eeqn

It is very useful to summarise FIC analyses
both in a table, with estimates $\hatt\mu_M$
of the focus parameter along with estimated
standard deviations and biases, and a FIC plot.
This plot displays the focus estimates $\hatt\mu_M$
on the vertical axis and $\fic_M^{1/2}$
in the horizontal axis, i.e.~the natural estimates
of the associated root-mse, transformed back
to the original scale of the focus parameter estimates.
Such FIC plots for the example on bird species 
are displayed in Section \ref{section:intro}.

Note that the methods exposited here are very
general, applicable for any regular parametric
set of models, with or without covariates,
as long as they are naturally nested between
well-defined narrow and wide models, and inside
a reasonable vicinity of each other. Also,
methods apply for any given focus parameter,
not merely for say those related to the mean responses.

\subsection{FIC for generalised linear models, via local asymptotics}
\label{subsection:fic-glm-local}

As pointed to above, FIC and AFIC formulae may be
derived in full generality for the class of
generalised linear models, using the local neighbourhood
models framework of Section \ref{subsection:fic-i};
see in this regard \citet{ClaeskensHjort08b}.
Here we are content to show how the apparatus
works for the class of logistic regression models.
The observations $y_i$ are hence 0 or 1,
with probabilities
\beqn
p_i=p(x_i,z_i,\beta,\gamma)
   =\Pr(Y_i=1\midd x_i,z_i)
   ={\exp(x_i^\tr\beta+z_i^\tr\gamma)
   \over 1+\exp(x_i^\tr\beta+z_i^\tr\gamma)}
   \quad {\rm for\ }i=1,\ldots,n.
\eeqn
This leads to a clear log-likelihood function
$\sumin \{y_i\log p_i+(1-y_i)\log(1-p_i)\}$,
to maximum likelihood estimators $(\hatt\beta_M,\hatt\gamma_M)$
in all submodels, with $M$ a subset of $\{1,\ldots,q\}$,
and in particular to submodel directed estimates
of any probability associated with a given individual,
say $\hatt p_M(x_0,z_0)=p(x_0,z_0,\hatt\beta_M,\hatt\gamma_M,0_{M^c})$
for an individual with covariates $(x_0,z_0)$.

To compute FIC scores, we start from the normalised
observed $(p+q)\times(p+q)$ Fisher information matrix,
which here becomes
\beqn
J_n=n^{-1}\sumin \hatt p_{i,\wide}(1-\hatt p_{i,\wide})
   \begin{pmatrix} x_i \\ z_i \end{pmatrix}
   \begin{pmatrix} x_i \\ z_i \end{pmatrix}^\tr,
\eeqn
with $\hatt p_{i,\wide}$ estimates of $p_i$ reached from
the wide model. Then we invert this to compute the
lower right-hand $q\times q$ block $Q_n$, along with
the relevant $\hatt G_{M,n}$ matrices described
in Section \ref{subsection:fic-i}. With the focus
being on a given individual, with linear predictor
$\mu=x_0^\tr\beta+z_0^\tr\gamma$, we have
$\omega=J_{n,10}J_{n,00}^{-1}x_0-z_0$, and may go on
to compute all relevant FIC scores, as with
(\ref{eq:ficone}), which here takes the form
\beqn
\fic^u(x_0,z_0)
   &=& n^{-1}\{\hatt\tau_0^2(u)
   +(z_0-\hatt J_{n,10}\hatt J_{n,00}^{-1}x_0)^\tr
    \hatt G_M\hatt Q\hatt G_M^\tr
   (z_0-\hatt J_{n,10}\hatt J_{n,00}^{-1}x_0) \\
& & \qquad
  +\,n(z_0-\hatt J_{n,10}\hatt J_{n,00}^{-1}x_0)^\tr (I_q-\hatt G_M)
   (\hatt\gamma_\wide\hatt\gamma_\wide^\tr-\hatt Q)
   (z_0-\hatt J_{n,10}\hatt J_{n,00}^{-1}x_0) \},
\eeqn
or its adjusted version with truncation to zero
of squared bias estimates.

The use of the R library {\tt fic} does not require 
that the user knows any of these formulae; it suffices to state 
the focus function, the narrow model, fit the wide model, 
and specify which of its submodels are of interest.

\subsection{Details and derivations for the fixed wide model FIC}
\label{subsection:detailsficwide}

In Sections \ref{subsection:fic-ii}--\ref{subsection:fic-iiii})
we saw how the mse for candidate estimators $\hatt\mu_M$ 
could be approximated and then estimated, in the setup 
with a fixed wide regression model. A crucial ingredient
in that development is the binormal approximation 
to the joint distribution of $(\hatt\mu_\wide,\hatt\mu_M)$,
formulated in statement (\ref{eq:approx3}). Here we give 
the details leading to that result. 
For this we need to go beyond the separate results 
(\ref{eq:approx1})--(\ref{eq:approx2}), related to limiting
distributions for model parameter estimators 
$\hatt\theta_\wide$ and $\hatt\theta_M$. Indeed, under 
general and mild regularity conditions, representations 
\beqn
\rootn(\hatt\theta_\wide-\theta_\true)
   &=&J_n^{-1}n^{-1/2}\sumin u(Y_i\midd x_i,\theta_\true)+o_\pr(1), \\ 
\rootn(\hatt\theta_M-\theta_{0,M,n})
   &=&J_{M,n}^{-1}n^{-1/2}\sumin u_M(Y_i\midd x_i,\theta_{0,M,n})+o_\pr(1)
\eeqn 
are in force, see e.g.~\citet[Appendix]{CLP16}. Featured here 
is the least false parameter $\theta_{0,M,n}$, defined as 
the minimiser of the Kullback--Leibler distance  
\beqn
\KL_n(f_\wide,f_M(\cdot,\theta_M))
   =n^{-1}\sumin\int f(y_i\midd x_i,\theta_\true)
   \log{f(y_i\midd x_i,\theta_\true)\over f_M(y_i\midd x_i,\theta_M)}\,\dd y_i. 
\eeqn 
From this, via the multi-dimensional Lindeberg theorem, 
and again under mild regularity, follows  
\beq
\label{eq:approxJ}
\begin{pmatrix}
\rootn(\hatt\theta_\wide-\theta_\true) \\
\rootn(\hatt\theta_M-\theta_{0,M,n})
\end{pmatrix}
%   \approx_d
%\begin{pmatrix}
%J_n^{-1}\rootn\bar U_n \\
%J_{M,n}^{-1}\rootn\bar U_{M,n}
%\end{pmatrix}
\approx_d\N_{p+p_M}(0,
\begin{pmatrix}
J_n^{-1} & J_n^{-1}C_{M,n}J_{M,n}^{-1} \\
J_{M,n}^{-1}C_{M,n}^\tr J_n^{-1} &J_{M,n}^{-1}K_{M,n}J_{M,n}^{-1}
\end{pmatrix}).
\eeq
This involves the $p\times p_M$ covariance matrix
\beqn
C_{M,n}=n^{-1}\sumin\E_\wide\,u(Y_i\midd x_i,\theta_\true)
   u_M(Y_i\midd x_i,\theta_{0,M,n})^\tr. 
\eeqn 
This properly generalises the separate results 
(\ref{eq:approx1})--(\ref{eq:approx2}), and leads 
via the delta method to (\ref{eq:approx3}). 

\section*{Acknowledgements}

Some of our FIC and AFIC calculations have been carried
out with the help of the R library {\tt fic}, developed by
Christopher Jackson, see {\tt github chjackson/fic}, 
also available on CRAN;
some of our algorithms are extensions of his.
C.C.~and N.L.H.~thank Kenji Konishi and the other scientists
at the Institute of Cetacean Research for obtaining
the body condition data for the Antarctic minke whale,
the IWC Scientific Committee's Data Availability Group
(DAG) for facilitating the access to these data,
and Lars Wall\o e for valuable discussions regarding
the modelling.
G.C.~acknowledges support of KU Leuven grant GOA/12/14, 
and C.C.~and N.L.H.~acknowledge partial support
from the Norwegian Research Council through
the FocuStat project at the Department of Mathematics,
University of Oslo.
Finally and crucially, the authors express gratitude
to three anonymous reviewers, for their detailed
suggestions, which led to a better and more clearly 
structured article. 

\bibliographystyle{frontiersinSCNS_ENG_HUMS}
% the particular style used by Frontiers
\bibliography{gerda_celine_nils2}

\end{document}